\documentclass[a4paper,notitlepage,abstract,10pt]{article} 
\usepackage{siunitx}
\usepackage{floatflt}
\usepackage{caption}
\usepackage{graphicx}
\usepackage[absolute]{textpos}
\usepackage{amsmath}
\usepackage{comment}
\usepackage[utf8]{inputenc}
\usepackage{lmodern}           
\usepackage[english]{babel}
\usepackage{tikz}
\usepackage{algorithmic}
\usepackage{algorithm}

\usepackage{wasysym}
\usepackage{amssymb}
\usepackage{listings}
\usepackage[final]{microtype} 
\usepackage{afterpage}
\usepackage{wrapfig}

\usepackage{color}
\definecolor{mygreen}{rgb}{0,0.6,0}
\definecolor{mygray}{rgb}{0.5,0.5,0.5}
\definecolor{mymauve}{rgb}{0.58,0,0.82}
\usepackage{csquotes}
\usepackage[style=authoryear, backend=bibtex]{biblatex}
\usepackage{hyperref}

\def\alignautorefname~#1\null{%
  Eq.~(#1)\null
}
\usepackage[subrefformat=parens]{subcaption}

\usetikzlibrary{automata,positioning}

\usepackage{authblk}
 \title{Parallel Statistical Multi-resolution Estimation}
\author[1]{Jan Lebert}
\author[1]{Lutz K\"unneke}
\author[2]{Johannes Hagemann}
\author[3]{Stephan C. Kramer\thanks{stephan.kramer@mpibpc.mpg.de}}
\affil[1]{University of G\"ottingen, Department of Physics, Friedrich-Hund-Platz 1, 37077, G\"ottingen}
\affil[2]{University of G\"ottingen, Institute of X-ray Physics, Friedrich-Hund-Platz 1, 37077, G\"ottingen}
\affil[3]{Max-Planck-Institut of Biophysical Chemistry,  Laboratory of Cellular Dynamics, Am Fa\ss{}berg 11, 37077 G\"ottingen}

\begin{document}
\newcommand{\Depth}{12.5}
\newcommand{\Height}{5}
\newcommand{\Width}{1}
\maketitle
\thispagestyle{empty}
\vspace{2cm}

\begin{abstract} 
\normalsize{ We discuss several strategies to implement Dykstra's projection algorithm on NVIDIA's compute unified device architecture (CUDA).
Dykstra's algorithm is the central step in and the computationally most expensive part of statistical multi-resolution methods. 
It projects a given vector onto the intersection of convex sets. 
Compared with a CPU implementation our CUDA implementation is one order of magnitude faster.
For a further speed up and to reduce memory consumption
  we have developed a new variant, which we call {\it incomplete Dykstra's algorithm}.
  Implemented in CUDA it is one order of magnitude faster than the CUDA implementation of the standard Dykstra algorithm.
  %

As sample application we discuss using 
the incomplete Dykstra's algorithm
as preprocessor for the recently developed super-resolution optical fluctuation imaging (SOFI) method \autocite{Dertinger29122009}.
We show that statistical multi-resolution estimation can enhance the resolution improvement of the plain SOFI algorithm just as the Fourier-reweighting of SOFI. 
The results are compared in terms of their power spectrum and their Fourier ring correlation \autocite{SaxtonBaumeister1982}.
The Fourier ring correlation indicates that the resolution for typical second order SOFI images can be improved
by about~30\%.

Our results show that a careful parallelization of Dykstra's algorithm enables its use in large-scale statistical multi-resolution analyses.

}
\end{abstract}

\newpage

\newpage

\section{Introduction}
\label{sec:intro}

An important topic in image analysis is the rejection of noise and blur from digital images.
A particular example is denoising and deblurring
 of micrographs in optical microscopy.
%
The unknown signal from object space is
convolved 
with the point 
spread function (PSF) of the imaging apparatus and disturbed by 
Poissonian noise. 
Digital image recording leads to a projection of the convolved signal onto a discrete set of points in space (pixels) and intensities, such that the micrograph is a vector of real data of finite length.
 The blur operator representing the PSF is ill-posed \autocite{vogel2002computational}
which makes its inversion in presence of noise a numerically expensive and difficult task.
Regularization techniques have to be applied, see for example \autocite{facciolo}. 
Although the usage of a regularization term stabilizes iterative reconstruction algorithms,
a regularization parameter has to be introduced, the choice of which is crucial for the estimator quality.
An inadequate choice of the regularization parameter leads to either a loss of  
details in the image, or to artifacts
from the ill-conditioned nature of the inversion problem. 
There are methods to choose the regularization in a spatially adaptive, iterative manner 
\autocite{gilboa, grasmair,hintermueller}. 
Spatially adaptive regularization methods detect areas of under- and over-regularization in an image
and then  locally adapt the regularization parameter.
Despite the success of these methods \autocite{review_adaptive}, 
the influence of the regularization parameter on the expected closeness of estimator to the true unknown signal 
is often not obvious. 

\begin{figure}[htb]
      \begin{center}
      \includegraphics[width=0.35\linewidth]{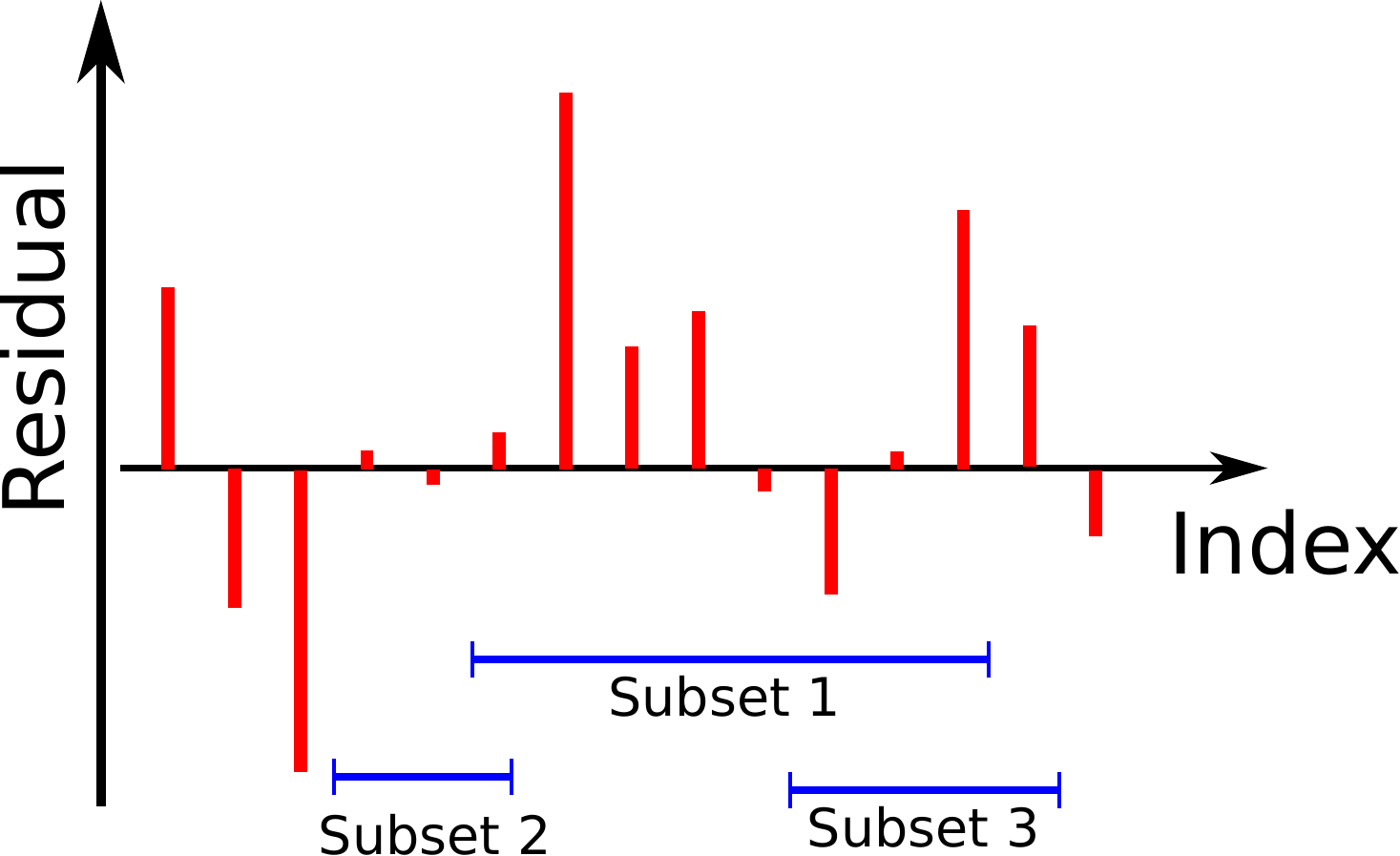}
      \hspace{1cm}
      \includegraphics[width=0.55\linewidth]{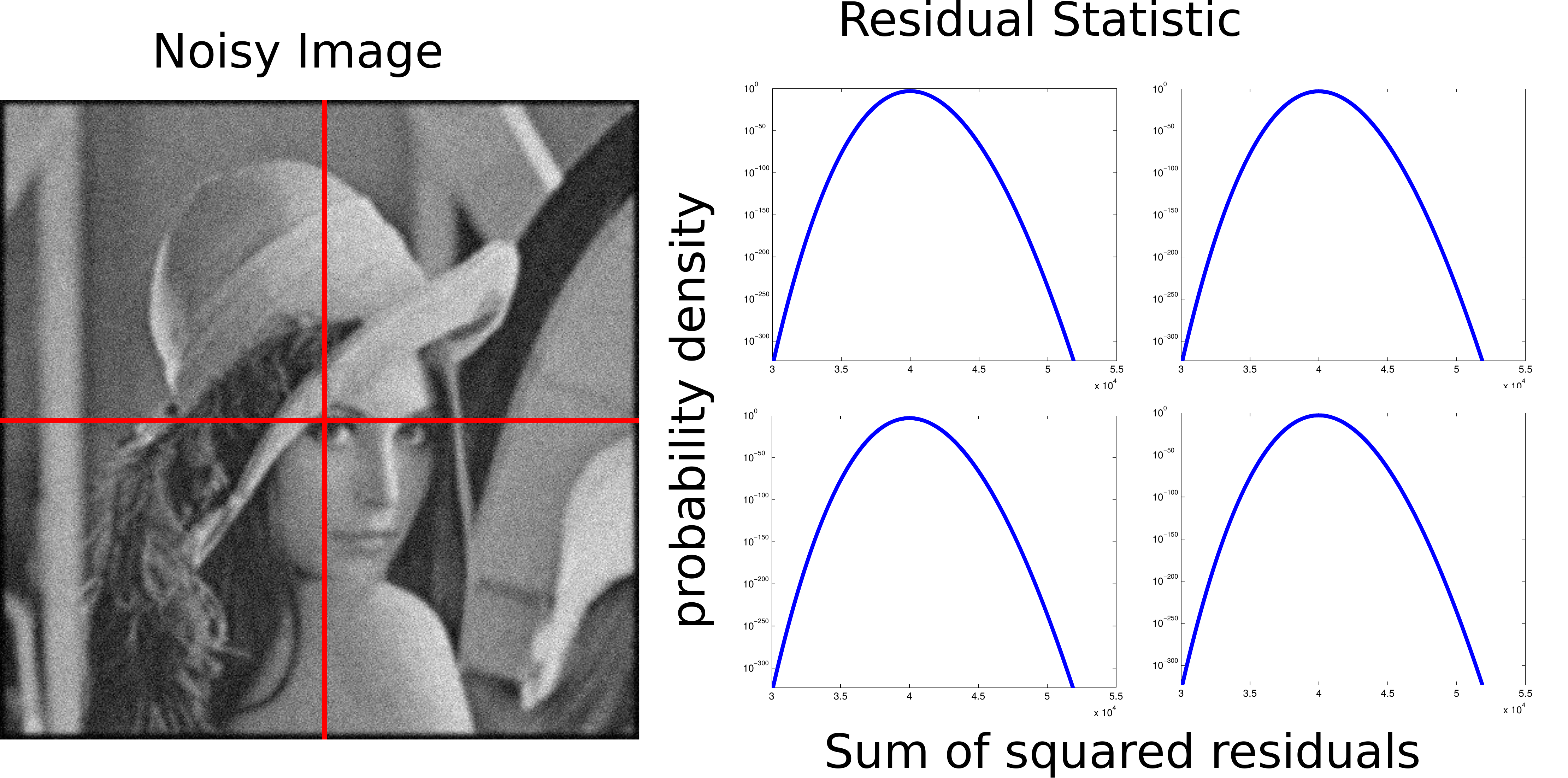}
       \caption{Left: The main concept of statistical multi-resolution estimation. 
       The residual statistics are considered in, possibly overlapping, subsets. The distribution of residuals 
       in each subset is required to be a realization of a given statistical process, i.e. Poisson or Gauss, 
       with chosen confidence $\alpha$. Right: Multi-resolution adapted to images. The subsets are parts of the image.
       The sum of squared residuals are required, in the example of normal distributed residuals,
       to follow a $\chi^2$ distribution.}
       \label{intro_fig}
      \end{center}
     \end{figure}

A slightly different approach to the problem is given by statistical multi-resolution
estimators (SMRE) \autocite{munk2, anscombe}. 
Common SMRE methods attempt to
    control the statistical behavior of residuals on several scales simultaneously, thus
     allowing the reconstruction of image details on different length scales at the same time.
     Figure \ref{intro_fig} illustrates the idea.
     The result is equivalent to a maximum-likelihood estimator with a properly chosen 
regularization parameter \autocite{chambolle}.
\par
The appealing feature of an SMRE is that the only free parameter in the resulting 
algorithm is the desired confidence level $\alpha$ in the hypothesis tests. 
Hence, the choice for the value of  $\alpha$ has 
a sound statistical interpretation.
\par
However, this comes at the price of an increased computational effort. 
The SMRE is to be computed iteratively. 
In each iteration
the residuals are to be controlled in each subset of the image, see for example one of the algorithms 
given in \autocite{anscombe}. 
\par
A common method to minimize a convex functional with respect to a set of independent variables
in presence of constraints is the alternating direction method of multipliers (ADMM) \autocite{admm_orig, admm}. 
One step in the ADMM is the projection of
    a given realization of residuals onto the allowed subset of residuals in
    $\mathbb{R}^n$. 
    Because of the nature of the constraints this is the projection 
    onto the intersection of many convex sets in Euclidean space. 
    Each set is given by one length scale. 
A solution to this task is given by
Dykstra's projection algorithm \autocite{dykstra_orig, dykstra}, which projects a given noise 
estimate onto the feasible set of the multi-resolution constraint.
A serial implementation of this algorithm is straightforward but
becomes infeasible as the number of considered scales increases.
\par
In recent years graphics cards have evolved from simple, dedicated graphics processing units (GPU) to autonomous parallel 
compute devices
within a computer, usually referred to as general-purpose
GPUs (GPGPUs).
With the introduction of CUDA (compute unified device architecture) and OpenCL there are powerful, yet sufficiently simple APIs (application programming interfaces) 
available for parallel computing.
\par
This work presents two different parallel implementations of Dykstra's algorithm optimized for NVidia's CUDA \autocite{Buck:2007:GCN:1281500.1281647}.
The first one implements Dykstra's algorithm in a general form.
The second implementation emerged from the work on the first one and introduces a variant which we call {\it incomplete Dykstra's algorithm} (ICD).
Its key feature is that only those subsets are chosen which fulfill the constraints of memory accesses on CUDA devices,
i.e. the subsets for a given scale have the shape of a tile, do not overlap, cover the whole image, the edge length is a power of two and the whole tile fits into the shared memory of one multiprocessor of a CUDA device. 
Despite its approximate nature, according to our tests presented in this paper, the ICD preserves the quality and statistical interpretation of the result obtained from the exact algorithm, but significantly improves the execution time.
The speedups of the CUDA implementations are measured on simulated test data.
The power of the new ICD is demonstrated by using it as preprocessor 
for the super-resolution optical fluctuation imaging (SOFI) method \autocite{Dertinger29122009}, a recent development in fluorescence microscopy.
As the number of frames for SOFI applications has to be large enough to resolve the statistical properties of the fluorescence signal, for instance in case of quantum dots up to several thousand images are required,
the performance of an SMRE is crucial for the feasibility of applying it as preprocessor.

This paper is organized as follows. Section \ref{sec:smre} contains a basic description of the theory of multi-resolution and introduces the ADMM, Dykstra's algorithm, SOFI and the Fourier ring correlation (FRC) for the quantitative assessment of the resolution improvement. 
Section \ref{sec:impl} highlights the key features of our implementation.
For a complete listing see the accompanying source code. 
In Sec.~\ref{sec:results}
we discuss our results, which include tests of the performance and the resolution improvement of the two variants of Dykstra's algorithm.  
To study the performance on real data obtained from an experiment we employ SMRE as pre- and postprocessor for SOFI. 
Finally, Sec.~\ref{sec:discussion} gives a conclusion.

\section{Statistical Multi-resolution Estimation}
\label{sec:smre}
In statistical signal processing the noisy measurement $I \in \mathbb{R}^m$ of the signal $x \in \mathbb{R}^n$  is formulated as  
\begin{align}
I = A*x^* + \epsilon \,,
\label{eq:basic1}
\end{align}
where $A$ is the PSF of the imaging apparatus, $x^*$ is the true signal underlying the measurement $I$ and $\epsilon$ is a noise vector. 
The convolution of~$A$ and~$x^*$ is denoted as $A * x^*$.
The estimates for the true signal and the noise, based upon knowledge of $I$, are written as $\hat{x}$ and $\hat{\epsilon}$, respectively.
In practice, both 
are unknown.
The linear operator $A$ is usually ill-posed which renders the inversion of Eq.~\eqref{eq:basic1} infeasible in any real world situation. 
In order to overcome this difficulty the problem is augmented with a regularization term $R \left( x \right)$.
A common example for the regularization is total variation (TV)
\begin{align*}
R_{\text{TV}} \left( x \right) = \|x\|_{\text{TV}} := \sum_{i=1}^{n} \sum_{k=1}^{d} \left|\left(\left(\nabla x\right)_i\right)_{k}\right|,
\end{align*}
where $d$ is 2 for planar images, and 3 for image stacks. The latter frequently occur in confocal microscopy.
The outer sum over $i$ is over all pixels and the inner sum computes the $L_1$ norm
of the local intensity gradient.
TV~estimators perform well on natural images \autocite{facciolo},
which are expected to consist of smooth areas and sharp boundaries.
The term $R$ may be substituted with a more problem-specific term, see for example \autocite{Diekmann2001526}.
\par
A multi-resolution estimator recovers the most regular vector 
\begin{align}
\hat{x} = \text{argmin}_{x, \epsilon}\; G \left( \epsilon \right) + R \left( x \right) \text{ s.t. } I = A * x + \epsilon 
\label{eq1}
\end{align}
with respect to $R \left( x \right)$, which fulfills the multi-resolution 
constraint 
\begin{align}
    G \left( \epsilon \right) = \begin{cases} 0  & \text{ if } \max_{s \in \Omega}
    c_s (\alpha) \sum_{i\in s} \epsilon_i^2 \leq 1 \\
\infty & \text{ else} \end{cases} \,.
\label{geq}
\end{align}

The function $G$ is a reformulation of the constraint that the noise $\epsilon$ has to be normally distributed, $\epsilon \sim N \left( 0 , \sigma^2 \right)$.
Using an Anscombe transformation \autocite{anscombe} 
the Poissonian noise of digital cameras can be transformed such that it approximately obeys a normal distribution. 
The image $I \in \mathbb{R}^m$ contains $m$ pixels, which are indexed by the index set $\{1 , ... ,m \}$.
A multi-resolution analysis works on subsets $s \subset \{1 , ... ,m \}$ of the pixel indexes. 
The set of all subsets of pixel indexes is denoted as  
\begin{align}
\Omega = \{ s ~ | ~s \subset \{1 , ... ,m \}\} \,.
\end{align} 
The feasible set generated by a subset $s$ of pixel indexes is denoted as
\begin{eqnarray}
\label{eq:tau_s}
\tau_s (x) := \{ x \in \mathbb{R}^n | c_s  (\alpha)\sum_{i \in s} x_i^2 \leq 1 \}\,. 
\end{eqnarray}
The intersection of all $\tau_s$ defines the feasible set
\begin{eqnarray}
\label{eq:tau_G}
\tau_G & := & \bigcap\limits_{s \in \Omega} \tau_s 
\end{eqnarray}
of the multi-resolution constraint $G$.

\begin{figure}[htb]
 \includegraphics[width=0.45\linewidth]{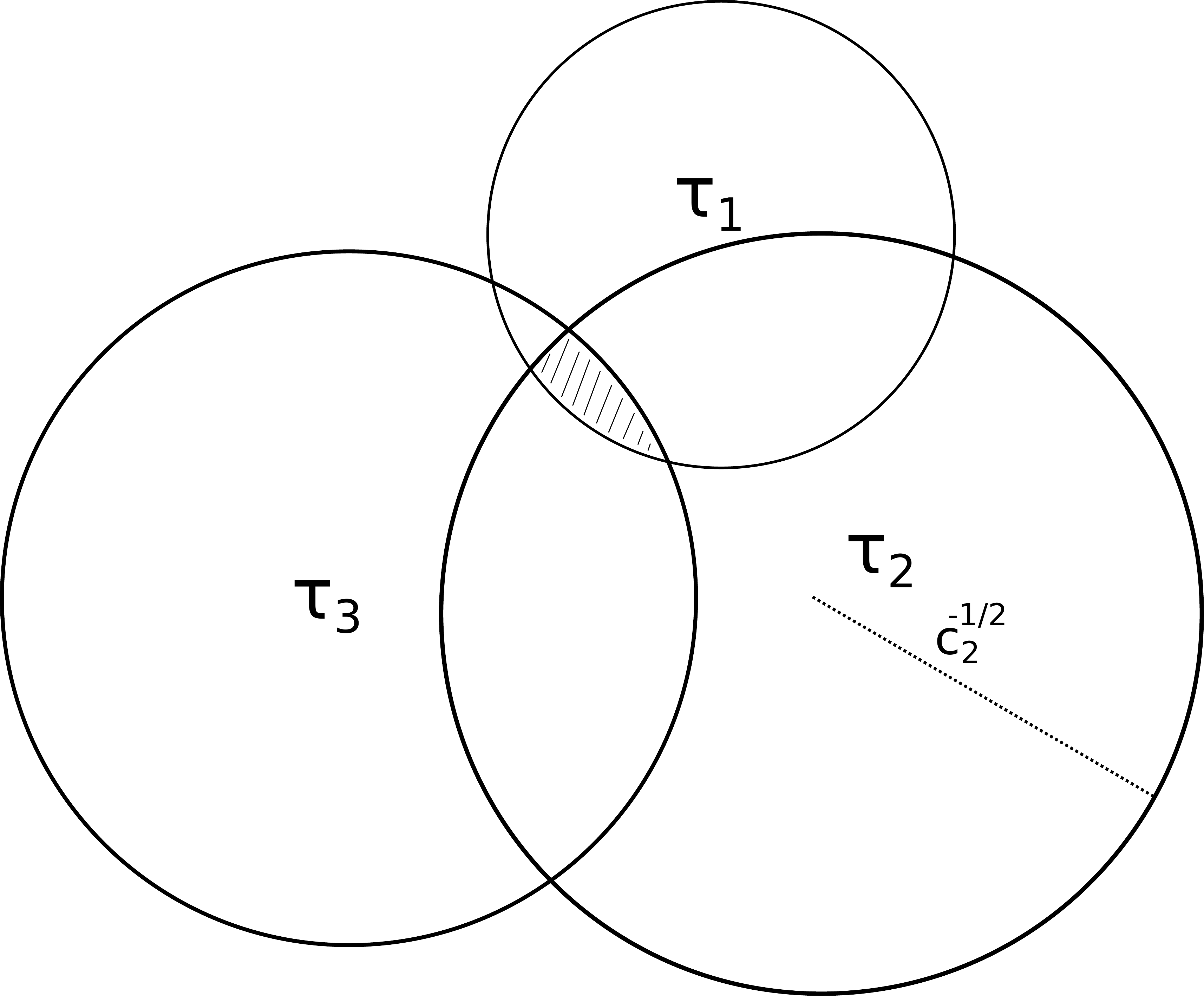}
 \includegraphics[width=0.45\linewidth]{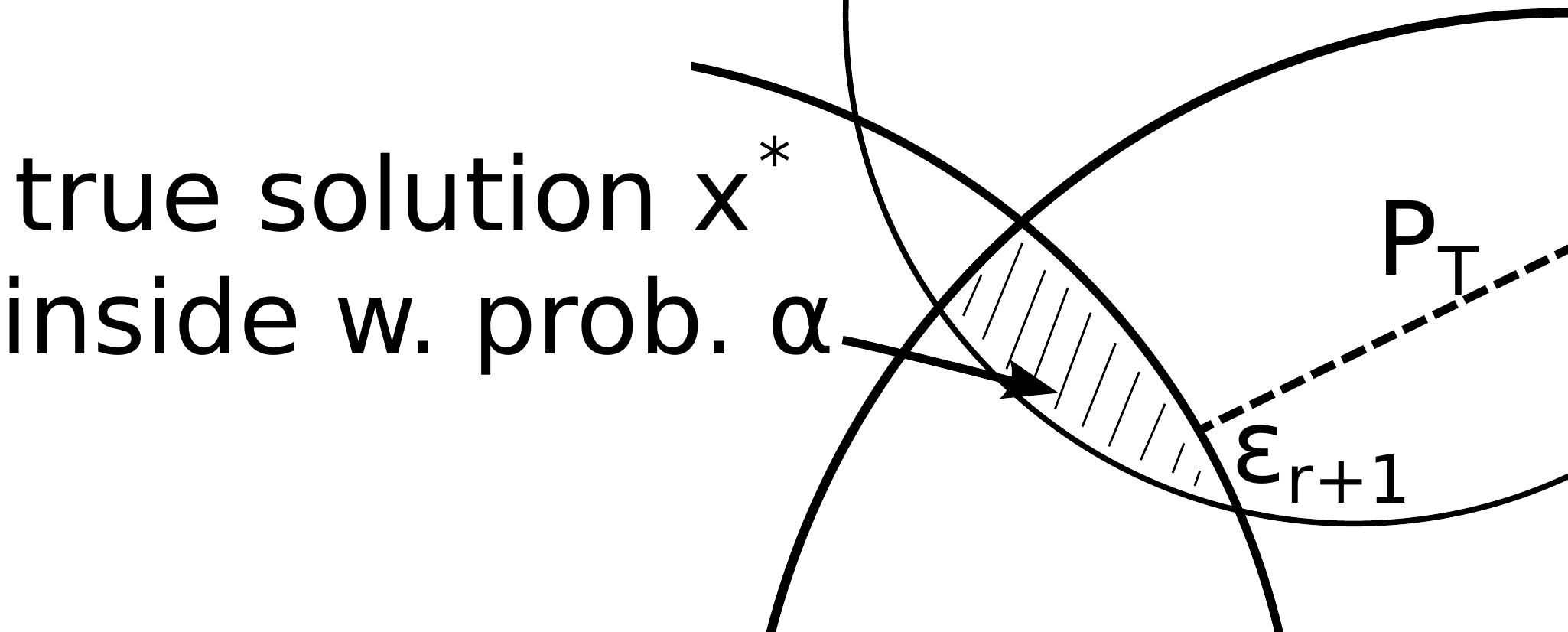}
 \caption{
 \textbf{Left:} The feasible sets are generated by the subsets $s \in \Omega$.  Their dimensions are given by the cardinalities of the subsets $s$. 
 The estimate $\hat{x}$ is inside of the intersection of all spheres $\tau_s$. 
 \textbf{Right:} The unknown true solution $x^*$ is inside of the feasible set with probability $\alpha$. }
 \label{circle_fig}
\end{figure}

In practical applications $\Omega$ will be restricted to connected areas of only a few pixels since the computational effort increases with $|\Omega|$, the cardinality of $\Omega$. 
Then $G$ computes the maximum squared residual on all subsets~$s \in \Omega$.
With an appropriate choice of the weights $c_s \left( \alpha \right)$ one can ensure that for each $e \in\{ \epsilon \in \mathbb{R}^m~|~G \left( \epsilon \right) < \infty \}$ 
 the noise $e|_{s}$ in the pixels belonging to a subset~$s  \in \Omega$ is Gaussian distributed with  a chosen probability of at least $\alpha$.
Hence, the estimator~$\hat{x}$ is the smoothest image in the sense of $R$
which satisfies the multi-resolution constraint enforced by $G$, as given in Eq.~\eqref{geq}.
This relation is visualized in Fig.~\ref{circle_fig}.
\subsection{Selection of the weights}
The weights $c_s(\alpha)$ are required to ensure a balanced hypothesis test among all chosen subsets $s \in \Omega$ of the image plane. 
The probability that $\max_{s \in \Omega} c_s (\alpha) \sum_{i \in s} \epsilon_s^2 \leq 1$ is required to equal $\alpha$ for all sets. 
A method to choose $c_s(\alpha)$ accordingly is given in \autocite{anscombe}. The function 
\begin{align*}
t_s \left( \epsilon \right) = \sum_{i \in s} \epsilon_i^2
\end{align*}
measures the sum of squared residuals in a set $s$.
Its fourth root transform 
\begin{align*}
\left( t_s \left( \epsilon \right) \right)^{\frac{1}{4}} \sim N \left( \mu_s = \left( | s | - 0.5 \right)^{\frac{1}{4}},\, \sigma_s^2=\frac{1}{8 \sqrt{|s|}} \right)
\end{align*}
is approximately normally distributed.
With this transformation each $s$ contributes equally to the extreme value statistics
\begin{align*}
Q_{\Omega} = \max_{s \in \Omega} \frac{ \left( t_s \right)^{\frac{1}{4}} - \mu_s}{\sigma_s} \,.
\end{align*}
Then, an appropriate choice for the weights is  
\begin{align*}
c_s(\alpha) = \frac{1}{\left( q_{\alpha} \sigma_s + \mu_s \right)^{4}}\,.
\end{align*}
In this equation $q_{\alpha}$ is the $\alpha$ quantile of $Q_{\Omega}$. 
It follows that 
$$P\left( \max_{s \in \Omega} c_s(\alpha) t_s
    \leq 1
\right) = \alpha\,, $$
i.e. the probability that Gaussian noise violates a single constraint is $\alpha$. 
The probability to violate the constraint is balanced for all subsets by the choice of the $c_s(\alpha)$.
Hence, the unknown true signal $x^*$ itself fulfills the constraint with probability $\alpha$,
\begin{align}
    P \left( G \left( I - A * x^* \right) \leq 1 \right) = \alpha \,.
\end{align}
Moreover, one can show \autocite{anscombe} a sound statistical relation between the estimated signal $\hat{x}$ and $x^*$
with respect to the regularization $R$ for the estimator in Eq.~\eqref{eq1} applied to Eq.~\eqref{eq:basic1},
\begin{align}
P \left( R \left( \hat{x} \right) \leq R \left( x^* \right) \right) \geq \alpha \,.
\end{align}
The probability that a feature in the recovered signal $\hat{x}$ is in fact a feature of the true image $x^*$ and not an artifact 
arising from incomplete noise removal is at least $\alpha$.

\subsection{Alternating Direction Method of Multipliers}
The numerical minimization of Eq.~\eqref{eq1} can be tackled by using the ADMM \autocite{admm_orig, admm}.
The ADMM is a method to minimize a convex functional depending on a set of variables in the presence of constraints. 
The constraints are enforced by augmenting the functional with Lagrangian terms $\Upsilon$, one for each constraint.
The convergence speed is enhanced by adding a quadratic penalty term $\| I - \left( A * x + \epsilon \right) \|_2^2$, weighted by a real number $\rho>0$ \autocite{admm}.
The constrained minimization problem in Eq.~\eqref{eq1} then corresponds to the unconstrained minimization 
with respect to the primal variables $x  \in \mathbb{R}^n$ and $\epsilon  \in \mathbb{R}^m$, and maximization
with respect to the dual variable $\Upsilon \in \mathbb{R}^m$ of the objective functional
\begin{eqnarray}
\mathcal{L} \left( x , \epsilon \right) &=& G \left( \epsilon \right) + R \left( x \right) + \frac{\rho}{2} \| I - \left( A * x + \epsilon \right) \|_2^2 \label{eq2} \\
& &+ ~\langle \Upsilon , I - \left( A * x + \epsilon \right) \rangle \nonumber \,.
\end{eqnarray}
The solution $\hat{x} = \text{argmin}_{x,\epsilon}\, \max_{\Upsilon}\, \mathcal{L}  \left( x , \epsilon \right)$
is obtained 
iteratively,
 cf.~\autoref{algo:admm}.
The estimate after $r$ iterations of the ADMM is denoted as $x_r$. In the limit $r\rightarrow \infty$ it converges towards the minimum,
$x_r \rightarrow \hat{x}$.
\begin{algorithm}
 \caption{Alternating Direction Method of Multipliers}
 \label{algo:admm}
\begin{algorithmic}
    \STATE choose $\delta > 0$, $x_0 \in \mathbb{R}^n$, $\Upsilon_0, \epsilon_0
    \in \mathbb{R}^m$, $\rho > 0 $
    \WHILE { $\| x_r - x_{r-1} \|_2 > \delta $ and $\| I - \left( A*x_r + \epsilon_r \right) \|_2 > \delta$ }
\STATE $x_{r+1} = \text{argmin}_x\, \mathcal{L} \left( x , \epsilon_r \right) $ \\
\STATE  $\epsilon_{r+1} = \text{argmin}_{\epsilon}\, \mathcal{L} \left( x_{r+1} , \epsilon
\right) $ \\
\STATE $\Upsilon_{r+1} = \Upsilon_r + \alpha \left( I - \left( A * x + \epsilon
\right) \right)$
\ENDWHILE
\end{algorithmic}
\end{algorithm}
The minimization with respect to $x$ can be evaluated in linear approximation, without damaging the overall convergence 
\autocite{admm}.
The update rule for $\Upsilon$ 
as given in \autoref{algo:admm} corresponds to the method of steepest ascent.
To simplify the minimization we decouple the deconvolution and smoothing by introducing another variable $z$. Hence 
the problem
\begin{align}
    \hat{x} = \text{argmin}_{x, \epsilon}\, G \left( \epsilon \right) + R \left( z \right) \text{ s.t. }   I = A *
    x + \epsilon \,, ~z = x\,,
    \label{eq3}
\end{align}
has two constraints.
This leads to the augmenting variables $\Upsilon_1$, $\Upsilon_2$, $\rho_1$, and $\rho_2$.
%
The result $\hat{x}$ is not altered by this additional constraint, 
but the individual steps in the ADMM become easier to solve. 
The introduction of $z$ does not perturb the convergence of the ADMM in linear approximation of $  \mathcal{L} \left( x , \epsilon_r \right) $.
To further reduce the error in each iteration step a stabilization term $\gamma \| x \|_2^2$ is introduced, which leads to a modified update rule
\begin{align*}
    x_{r+1} = \text{argmin}_{x} &\, \frac{\rho_1}{2} \| I - A*x - \epsilon \|_2^2 +
    \langle \Upsilon_1 , I - A*x - \epsilon \rangle + \\ &+ \frac{\rho_2}{2} \| x - z
    \|_2^2  + \langle \Upsilon_2 , x - z \rangle 
\end{align*}
which can be used in its linearized version
    \begin{eqnarray}
   x_{r+1} & \approx & \text{argmin}_{x} \, \gamma \|x\|_2 + \langle \Upsilon_2 - A^T * \left(
    \rho_1 \left(
    I - \epsilon \right) + \Upsilon_1 \right) - \rho_2 z , x \rangle \,.
\end{eqnarray}
Here, $A^T$ denotes the transpose of the matrix $A$.
The minimization with respect to $x$ has no closed form since $A$ is ill-posed.
As long as $\gamma$ is chosen sufficiently large, $x_r$ still converges to $\hat{x}$ for large $r$.
The minimization with respect to $\epsilon$ can be rewritten as a projection on the intersection of convex sets,
\begin{eqnarray}
\epsilon_{r+1} &= &
 P_G \left( I - A*x + \frac{\Upsilon}{\rho} \right)\,,
\end{eqnarray}
where, $P_G(\cdot)$ is defined as
\begin{eqnarray}
P_G(X) & := &  \text{argmin}_{\epsilon} \left\lbrace G(\epsilon) +  \frac{\rho}{2} \left\| \epsilon - X \right\|_2^2 \right\rbrace \,.
\end{eqnarray}
%


\subsection{Dykstra's Algorithm}
\label{sec:intro:dykstra}
The problem of projecting on the feasible set $\tau_G$ of the multi-resolution constraint $G$ is solved using Dykstra's algorithm \autocite{dykstra_orig,dykstra}, which requires
the knowledge of 
the projections $p_s$ on each of the
sets $\tau_s (\epsilon)$.
In general, the projection on a set $\tau_s (x) $
 is defined as
\begin{align}
    p_s \left( x_{i\in s} \right) = 
    \begin{cases} \frac{x_i}
    {
    \sqrt{c_s  (\alpha) } 
    \| x \|_{2, s}
    }  
     & \text{ if } c_s  (\alpha)  \sum_{i \in s} x_i^2 > 1 \\ 
        x_i & \text{ else} \end{cases} 
        \,,
         \label{eq:dykstra:if}
\end{align}
where $ \| x \|_{2,s}^2 :=  \sum_{i \in s} x_i^2 $ is the $L_2$ norm with respect to the pixels belonging to the set~$s \in \Omega$.
This projection is to be performed in
each iteration of the ADMM. Therefore, its performance is critical for the overall minimization. 
The projection on $\tau_G$ 
 is computed iteratively as shown in \autoref{algo:dykstra}.
\begin{algorithm}
 \caption{Dykstra's Algorithm}
 \label{algo:dykstra}
\begin{algorithmic}
    \STATE choose $\delta > 0$, given $x_0, q_0 \in \mathbb{R}^n$ with $q_{0,s} =
    0 \, \hspace{0.5cm} \forall s \in \Omega$
    \WHILE { $\| x_r - x_{r-1} \|_2^2 > \delta $ }
\STATE $x_{r+1} = p_s \left( x_r - q_{r,s} \right) $
\STATE $q_{r+1,s} = x_{r+1} - x_r $
    \ENDWHILE
\end{algorithmic}
\end{algorithm}
The algorithm is known to converge towards the projection of~$x_0$ on the intersection of the $s \in \Omega$. 
The iteration is stopped when the change in~$x_r$ is below some chosen threshold $\delta$. 
In the context of image processing the sets~$s$ denote 
small subsets in the image. 
The projection on two sets $s_1, s_2 \in \Omega$ can be 
calculated in parallel if $s_1 \cap s_2 = \emptyset$. However, the order in which the projections are 
performed is crucial for the convergence of the algorithm.  Therefore, only non-overlapping subsequent projections 
can be calculated in parallel.
\subsection{SOFI}
\label{section:sofi}
Recently, several methods have been developed to
overcome the diffraction limit (first derived by Abbe) on resolution in optical microscopy,
such as PALM (photo-activation localization microscopy) \autocite{Hess20064258},
 STORM (stochastic optical reconstruction microscopy (STORM) \autocite{Rust:2006fk}
 SIM (structured illumination microscopy) and its nonlinear variant \autocite{Gustafsson13092005} and most notably the STED (stimulated emission depletion) \autocite{Hell:94} microscopy which was awarded the Nobel prize in Chemistry, 2014.
 All of these methods require to some extent special sample preparation techniques making these methods more elaborate and complicated than traditional fluorescence wide-field microscopy.

 A different approach is SOFI (super-resolution optical fluctuation imaging) \autocite{Dertinger29122009, 2010OExpr1818875D} which infers the additional information necessary for resolution improvement from 
  the temporal behavior of the signal of the imaged object
which in 
 case of fluorescence microscopy works as follows.
Given $N$ point-like emitters, the fluorescence signal at an arbitrary point ${\bf r} \in \mathbb{R}^3$ (in practice this is a pixel) in the image plane can be written as
\begin{eqnarray}
\label{eq:fluo-sig}
F({\bf r}, t) & = & \sum_k U({\bf r} - {\bf r}_k) \varepsilon_k s_k(t) \,,
\end{eqnarray}
where $U :  \mathbb{R}^d \to  \mathbb{R}^d$ is the PSF, which is entirely determined by the optical system and time-independent.
In the following we restrict the discussion to the focus and image plane, i.e. $d=2$.
The molecular brightness of the $k$th fluorophore is denoted as $\varepsilon_k \in \mathbb{R}^+$ and the stochastic time-dependence of its emitted fluorescence is $ s_k(t) : \mathbb{R} \to [0,1]$.
With $\langle \ldots \rangle_t$ as average over time,
the fluctuations $\delta F({\bf r}, t) = F({\bf r}, t) - \langle F({\bf r}, t)  \rangle_t$ of the observed fluorescence are given by
\begin{eqnarray}
\label{eq:fluo-sig-fluc}
\delta F({\bf r}, t)  & = & \sum_k U({\bf r} - {\bf r}_k) \varepsilon_k \delta s_k(t) \,.
\end{eqnarray}
To motivate the idea underlying the SOFI signal we consider the two-point correlation function $G_2({\bf r}, \tau ) = \langle \delta F({\bf r}, t + \tau ) \cdot \delta  F({\bf r}, t)  \rangle_t $ which by Eq.~(\ref{eq:fluo-sig}) is 
\begin{eqnarray}
\nonumber \\
\label{eq:sofi-G2}
G_2({\bf r}, \tau ) & = & \sum_k U^2({\bf r} - {\bf r}_k) \varepsilon^2_k \langle \delta s_k(t + \tau ) \cdot \delta  s_k( t)  \rangle_t \,.
\end{eqnarray}
To get a value equivalent to the intensity the final SOFI signal is given by the integrated correlation
\begin{eqnarray}
I_{SOFI} ({\bf r}) 
\label{eq:sofi-sig}
& = & \sum_k U^2({\bf r} - {\bf r}_k) \varepsilon^2_k \int\limits_{-\infty}^{+\infty}
\langle \delta s_k(t + \tau) \cdot \delta s_k( t) \rangle_t ~d \tau \,. 
\end{eqnarray}
In general, instead of correlation functions cumulants $C_n$ are used, where $n$ denotes the order. The advantage of cumulants is the absence of correlations of orders less than~$n$.
Then, the intensity $I_{SOFI} ({\bf r})  $ assigned to a pixel in the final SOFI image is proportional to the $n$th power of the PSF. Usually, a Gaussian profile is a good approximation to the shape of the PSF and thus resolution is improved by a factor of $\sqrt{n}$ if a cumulant of order $n$ is used.
However, at the same time inhomogeneities in the molecular brightnesses $\varepsilon_k$ are amplified as well, as already the simple case of order~2 in Eq.~(\ref{eq:sofi-sig}) shows.

A common simplification, especially for higher order SOFI methods, is to approximate the integral by the maximal value of the integrand, which for order 2 is simply the variance of the intensity in a pixel 
\begin{eqnarray}
\label{eq:sofi-sig-var}
C_2({\bf r}, 0) & = & \langle F^2({\bf r}, t ) \rangle_t - \langle F({\bf r}, t ) \rangle_t^2\,.
\end{eqnarray}
%

For higher orders the cumulants can be computed from the moments of the probability distribution of the fluctuations.

The strength of SOFI is that it can be combined with a variety of microscopy techniques without any need for modifying the experimental setup as it is a software-only, pure post-processing method.
Its disadvantage is that it needs a sufficient amount of images to resolve the statistical behavior of the temporal fluctuations of the observed (usually fluorescence) signal.
However, STORM has the same drawback.

\subsection{Fourier ring correlation}
\label{sec:frc}
In order to assess the quality of an estimator on experimental data an estimate for image resolution is necessary.
The  Fourier ring correlation (FRC) quantifies the image resolution from correlations in Fourier space and the impact of noise on those correlations.
It was specifically designed for the case where the true signal is not known.
Originally, the FRC was conceived as resolution measure for electron microscopy images \autocite{SaxtonBaumeister1982}.
 Recently, it was proposed as resolution criterion for optical super-resolution
microscopy \autocite{frc2013-2,frc2013}. 

To compute an FRC 
the correlation along circles in the Fourier plane is calculated for different realizations of the noise. 
The FRC defines  
image resolution as the smallest distance at which the correlation of two considered realisations 
of the same statistical process drops below the predicted correlation for pure noise.
Given two realizations $I_1, I_2 \in \mathbb{R}^n$ of the same noisy image acquisition process,
 the correlation 
of the Fourier transformed images $\widetilde{I_1}$, $\widetilde{I_2}$, constrained to a ring in distance $r$ to the origin, is calculated as
\begin{align}
 FRC(r) = \frac{\sum_{r_i \in r}{ \widetilde{I_1}(r_i) \cdot \widetilde{I_2^{\ast}}(r_i) }}
{\sqrt{\left(\sum_{r_i \in r}{ \left|\widetilde{I_1}(r_i)\right|^2}\right) \cdot \left(\sum_{r_i \in r}{\left|\widetilde{I_2}(r_i)\right|^2}\right)}}.
\end{align}

\begin{figure}[htpb]
 \includegraphics[width=\textwidth]{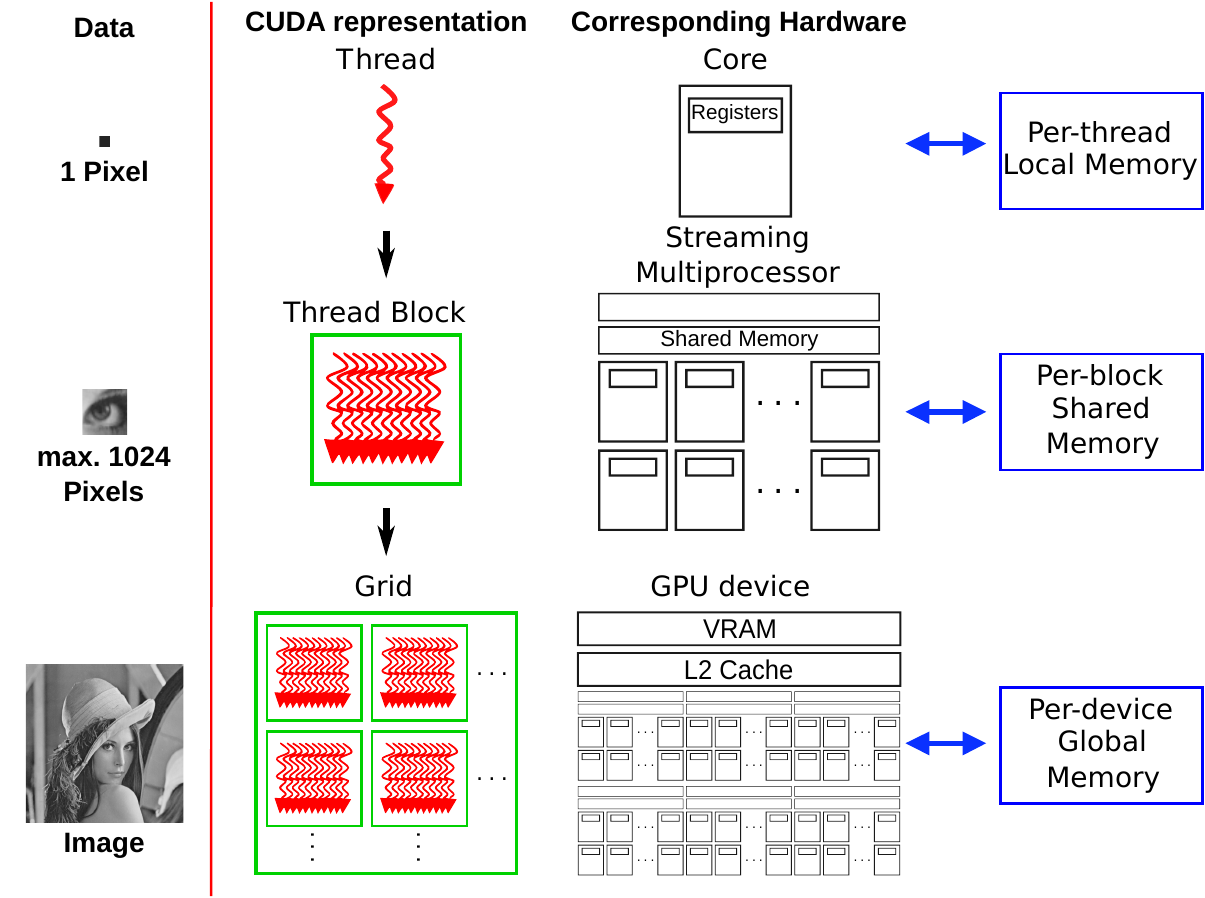}
\caption{The block diagram illustrates the mapping of parallel processing of data (left) to the programming structure of CUDA (middle). This structure is mainly modeled following the hardware design of GPUs (right), especially the memory hierarchy.}
\label{fig:cuda}
\end{figure}

\section{Implementation}
\label{sec:impl}
The core of an implementation of an SMRE is the ADMM algorithm for minimizing Eq.~\eqref{eq3}. 
The operator $A$ is approximated as Gaussian kernel with the standard deviation $\sigma_{\textrm{PSF}}$ given as input parameter, cf. the parameter listing in the appendix.
The implementation is based on the SciPAL library. 


The numerically expensive part is the implementation of Dykstra's algorithm, cf. \autoref{algo:dykstra}.
We first discuss the implementation which strictly follows the mathematical description as given in Sec.~\ref{sec:intro:dykstra},
which we call the \emph{exact method} (Sec.~\ref{sec:impl:exact}).
Then we discuss our \emph{incomplete method} (Sec.~\ref{sec:impl:incomplete}), which, according to our numerical experiments, is equivalent to the exact method,
but much simpler to implement because it partitions the image in non-overlapping sets right from the beginning. 
For both variants the runtime mainly depends on the maximum size of the subsets and the size of the image.
In general, subsets and images may have virtually any shape. Therefore, we introduce an effective edge length $L_s$ for the subsets and $L_I$ for images. Both are defined as the square root of the number of pixels in a subset and an image, respectively.
We further define the maximum multi-resolution depth
\begin{eqnarray}
\label{eq:k-smre}
K_{\text{SMRE}} & := & \log_2 \left( \max_sL_s\right) 
\end{eqnarray}
and the associated maximum resolution scale
\begin{eqnarray}
L_{\text{SMRE}} & := & 2^{K_{\text{SMRE}} } \,.
\end{eqnarray}
For readers not familiar with the CUDA programming model we provide a brief description in Appendix~\ref{sec:appendix-cuda-arch}. 
A graphical summary of our way mapping data to the GPU  is given in Fig. \ref{fig:cuda}.

\subsection{Dykstra's Algorithm (exact Version)}
\label{sec:impl:exact}

\begin{figure}[p]
\centering
\begin{minipage}{0.4\textwidth}
\centering
  \begin{subfigure}{\linewidth}
    \includegraphics[width=0.85\linewidth]{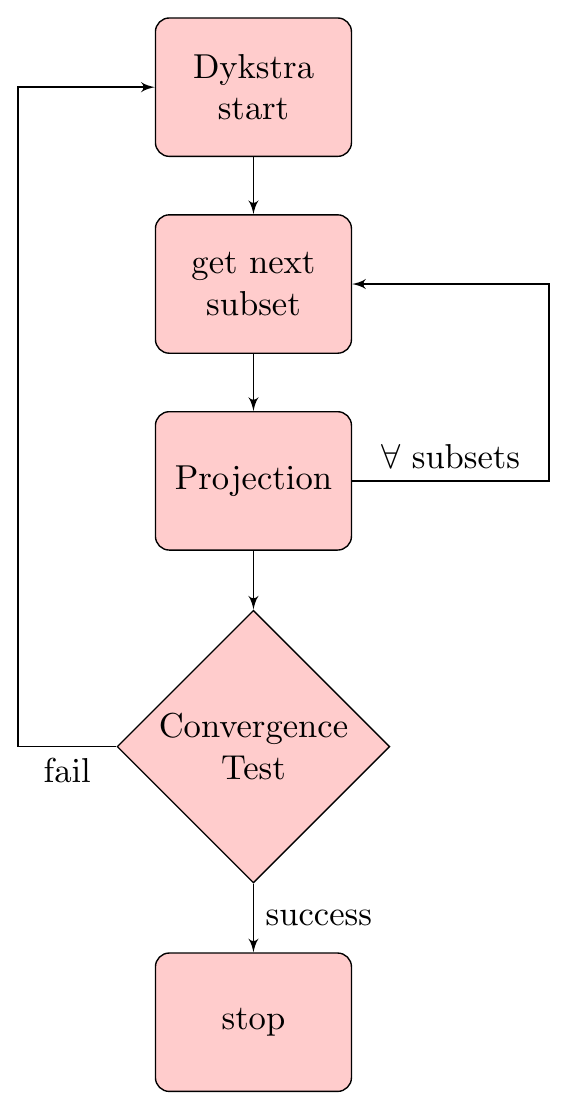}
    \caption{serial}
    \label{fig:flow:serial}
  \end{subfigure}\\[\baselineskip]
  \begin{subfigure}{\linewidth}
    \includegraphics[width=0.85\linewidth]{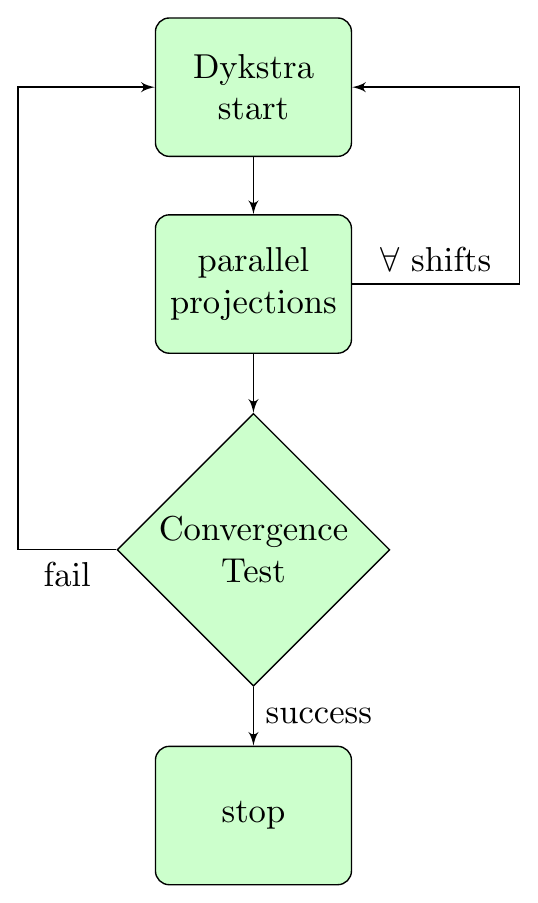}
    \caption{incomplete}
    \label{fig:flow:incomplete}
  \end{subfigure}
\end{minipage}
\hfill
\begin{subfigure}{0.5\textwidth}
  \includegraphics[width=0.85\textwidth]{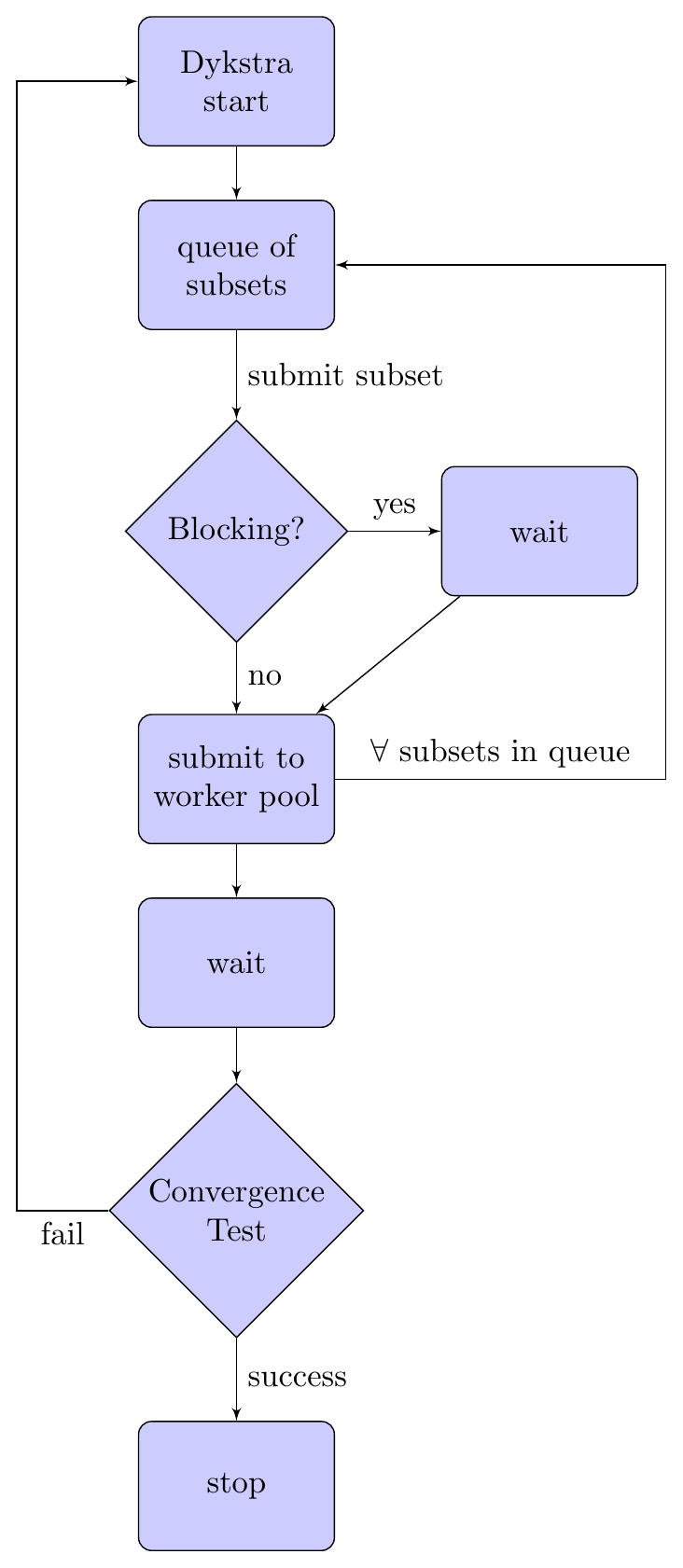}
  \caption{exact}
  \label{fig:flow:exact}
\end{subfigure}
\caption{Differences between a serial implementation \subref{fig:flow:serial} and the presented parallel 
exact \subref{fig:flow:exact} and incomplete \subref{fig:flow:incomplete} implementations. 
Only non-overlapping subsequent projections can be calculated in parallel and the sequence of overlapping subsets needs to be preserved.
This constraint shaped the implementation of the exact method, as it has to check for overlapping (blocking) subsets. 
The performance of the exact method is mainly limited by the transfer rate of the PCIe bus due to the limited RAM on the GPU.
See Fig.~\ref{fig:kernels} for a detailed discussion on the CUDA kernels of the parallel projection methods. 
The incomplete method works on a set of subsets which is tailored for concurrent execution. 
This increases the performance. 
The serial approach performs the projections one after another.}
\label{fig:flow}
\end{figure}
The parallelized implementation of Dykstras's algorithm is the key to reach high speedups.
Figure \ref{fig:flow} shows simplified block diagrams of the serial implementation~\subref{fig:flow:serial}, of the ICD \subref{fig:flow:incomplete} and of the exact method~\subref{fig:flow:exact}.
Only non-overlapping subsequent projections $p_s$ can be calculated in parallel and the sequence of overlapping subsets needs to be preserved.
This constraint is the principal guiding line for the implementation. 

\par
%
Each group is limited to a maximum size of 1024 pixels per group if each thread works on one pixel as this is the maximum number of threads per thread block. 
If the next set to be added to a group during creation exceeds this limit a new group is formed, leaving some threads idle in the former group. Therefore, in general not all threads of a thread block are in use.
To avoid blocking behavior within a group, the subsets of pixels must be chosen such that   
 only projections on mutually non-overlapping sets occur.
The groups of subsets are stored in a linked list, in the following called execution queue, which defines the sequence in which they are processed.
The order in which projections are carried out has to be such that the distance between overlapping subsets is large. 
In this context, distance is meant with respect to the positions in the execution queue.

\par
The efficient computation on GPUs faces two further problems.
(i) The limited amount of VRAM is not sufficient to store all necessary data on the GPU.
The memory needed to store all $q_s$ variables crucially depends on the multi-resolution scale $L_{\text{SMRE}} $ and the number of pixels in an image $L_I^2$, and  is $\mathcal{O}(L_{\text{SMRE}} ^3 L_I^2)$.
For instance, for a $1024\times1024$ pixel image and $L_{\text{SMRE}} =15$ this amounts to \SI{14}{GB} for single-precision.
(ii) To populate all the SMs in order to achieve good efficiency when calculating the parallel projections.
The classical approach of starting a single CUDA kernel is not efficient due to the blocking behavior of overlapping subsequent projections 
and the insufficient amount of memory on the GPU.
We have to use the conventional RAM on the CPU side as additional buffer for temporary data which requires frequent memory transfers before and after the execution. 
Instead, we make use of CUDA's Hyper-Q feature to implement a multi-stream based parallelization. 
\par
A key element of our implementation is a pool of worker threads on the host side, where each one uses a CUDA stream to process items received from a shared queue.
The queue holds the groups of subsets. 
Before adding the groups to the queue, the main worker thread 
verifies that the group to be added does not overlap with groups already queued. 
If an overlap is detected, the thread waits until the blocking group is processed and
removed from the queue. 
Only the main thread which adds groups of subsets to the queue is subject to long blocking behavior.
The GPU always has a high work load as memory transfers and instruction execution are done concurrently on multiple streams. 
\autoref{lst:exact} shows the stream handling in the worker thread which coordinates the work on a CUDA stream.
It retrieves several groups of subsets at once from the queue, and then 
adds copy and execution instructions to a CUDA stream to process a list of groups (a cluster) in parallel.
For brevity, uninteresting parts of the source code like error checking have been omitted from the listing.
Groups are designed to be processed by a thread block. 
This implementation launches the kernel in a grid of many thread blocks. 
The memory transfers are sufficiently large to make the use of multiple streams an effective latency hiding mechanism. 

\begin{lstlisting}[language=c++, mathescape, morekeywords={__global__, __device__, __shared__, threadIdx}, 
escapebegin=\color{mygreen}, caption={The parallel worker threads which control the CUDA streams}, label=lst:exact]
// This function is executed by `stream_count` threads in 
// parallel, each creates an CUDA stream and coordinates 
// the work done on that stream. It removes items from 
// the queue and processes them. End thread when terminate
// item is recieved from the queue.

// `group` is a class which contains a group of subsets
// `T` can be float or double
template<typename group, typename T>
void stream_handler() {
    // CUDA device id has to be set by each thread
    cudaSetDevice(inf->device_id);
    
    // Create CUDA stream
    cudaStream_t mystream;
    cudaStreamCreate(&mystream);

    // Remove several groups from queue on each get() call
    std::list<group*> cluster_vec;
    // Maximum number of groups we try to get from queue
    const int max_num_cluster = 128;
    // Minimum number of groups we try to get from queue, not guaranteed
    const int min_num_cluster = 12;

    // Allocate the maximum needed memory on host and device for
    // cluster information and q_offset

    // Array of device pointers which contains the information about the clusters
    
    SciPAL::Vector<*int, blas> cluster_info(4*max_num_cluster);
    SciPAL::Vector<*int, cublas> cluster_info_d(4*max_num_cluster);

    // Where to find the $q$ values for each thread block
	SciPAL::Vector<int, blas> q_offset(max_num_cluster);
	SciPAL::Vector<int, cublas> q_offset_d(max_num_cluster);
	
    // All $q$ values
	SciPAL::Vector<T, cublas> q_d(1024*max_num_cluster);

    step35::Kernels<T> kernels;

    // Get a list of min_num_cluster to max_num_cluster clusters
    cluster_vec = queue->get(min_num_cluster, max_num_cluster);

    // Big while loop until terminate signal is received
    while ( cluster_vec.front()->size != 0 ) {
        // Number of groups we received from queue
        int num_of_cluster = cluster_vec.size();

        // Fill arrays q_offset and cluster_info with the 
        // device pointers to the desired information
        // ...
        
        // In the big $q$ array, where can I find the $q$ for 
        // the first pixel of my cluster?
        unsigned int offset = 0;
        
        // Copy $q$ for all frames to device
        for (auto it=cluster_vec.begin(), 
             end=cluster_vec.end(); it!=end; ++it) {
            cudaMemcpyAsync(&(q_d[offset]), (*it)->qmat, 
                            (*it)->size*sizeof(T),
                            cudaMemcpyHostToDevice, mystream);
            offset+=(*it)->size;
        }

        // Add copy command of cluster_info and q_offset 
        // from host to device to CUDA stream
        // ...
        
        // Start a Dykstra's CUDA kernel on all frames in 
        // the list of clusters we received
        kernels.dykstra(q_d, inf->e_d, cluster_info_d, q_offset_d,
                        num_of_cluster, inf->width, inf->height,
                        1024, &mystream);

        // Add copy command of $q$ from device to host to
        // our CUDA stream
        // ...

        // Block until our CUDA stream has completed all operations
        cudaStreamSynchronize(mystream);
        
        // Signal queue that task is done
        queue->task_done(cluster_vec);

        // Get a list of min_num_cluster to max_num_cluster
        // groups from queue
        cluster_vec = queue->get(min_num_cluster, max_num_cluster);
    } // End big while loop

    // Worker thread shutdown cleanup
    // ...
}
\end{lstlisting}

\subsection{Incomplete Projection} 
\label{sec:impl:incomplete}

\begin{figure}
\centering
\hfill
\begin{subfigure}{0.45\textwidth}
 \includegraphics[width=\textwidth]{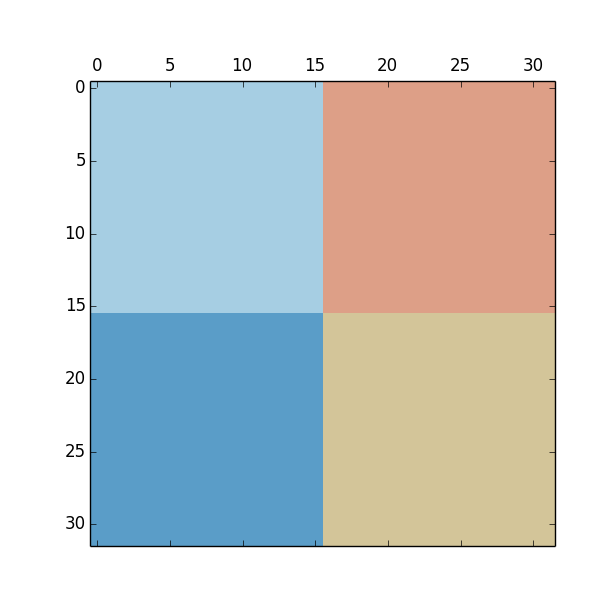}
 \caption{Thread block pattern used in both incomplete and exact method}
 \label{fig:algo:s4}
\end{subfigure}
\hfill
\begin{subfigure}{0.45\textwidth}
 \includegraphics[width=\textwidth]{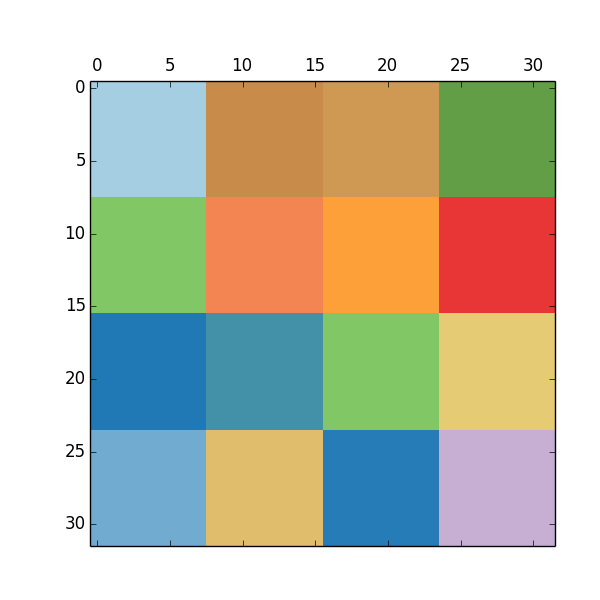}
 \caption{Thread block pattern used in both incomplete and exact method}
 \label{fig:algo:s3}
\end{subfigure}
\hfill\vfill\hfill
\begin{subfigure}{0.45\textwidth}
 \includegraphics[width=\textwidth]{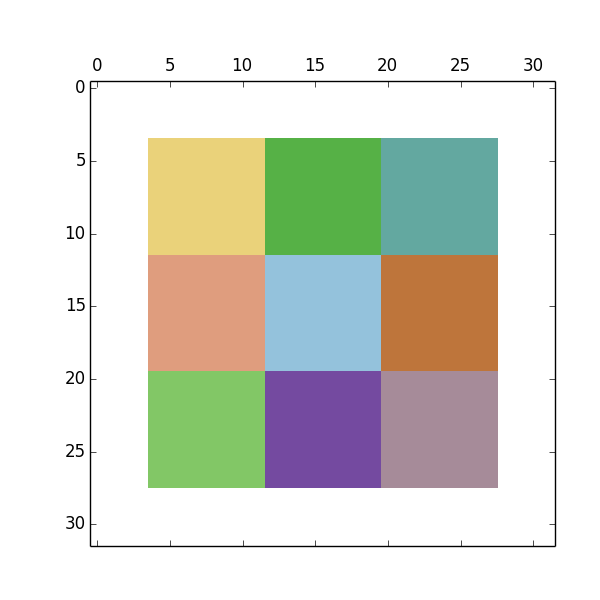}
 \caption{Thread block pattern used only in exact method}
 \label{fig:algo:frame415}
\end{subfigure}
\hfill
\begin{subfigure}{0.45\textwidth}
 \includegraphics[width=\textwidth]{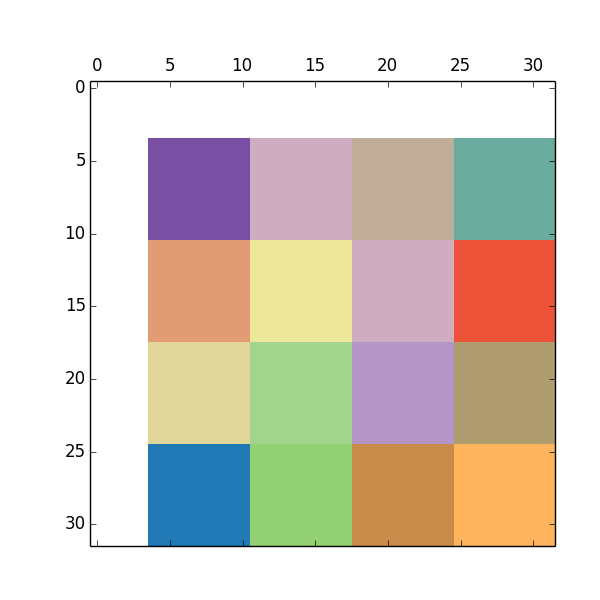}
 \caption{Thread block pattern used only in exact method}
 \label{fig:algo:frame388}
\end{subfigure}
\caption{Visualization of the differences between the exact and the incomplete method. The illustrations show $32 \times 32$ pixel sections of the larger image, 
each section is processed by one CUDA thread block of 1024 threads. 
%
The incomplete method makes optimal use of thread blocks by using subsets of $\Omega$ of edge length $1, 2, 4, 8, 16$ and $32$ pixels which fill up a $32 \times 32$ pixel section.
The subfigures \subref{fig:algo:s4} and \subref{fig:algo:s3} show the pattern of subsets of edge length $16$ and $4$ pixels respectively.
While the exact method uses all subset patterns of the incomplete method it also includes subsets of $\Omega$ with all edge lengths and all possible shifts within the image.
The patterns \subref{fig:algo:frame415} and \subref{fig:algo:frame388} are thus only used in the exact method and not in the incomplete method.
}
\label{fig:kernels}
\end{figure}

The exact variant has some limitations which hinder an 
efficient CUDA implementation, such as the PCIe bottleneck and the overall complexity of the detection of overlapping sets.
To further adapt Dykstra's algorithm to the execution on a CUDA device we now restrict the eligible subsets of the image plane to only those which are formed by squares with edge
lengths of powers of 2, up to $L_{\text{SMRE}} = 32$. 
Fig.~\ref{fig:flow} shows a general overview of the discussed implementations, Fig.~\ref{fig:kernels} shows the difference in
the chosen subsets between the exact and incomplete implementation.

All shifts which are a power of 2 up to 32 of these squares are considered. 
Consider the set of non-overlapping squares  in the image plane with edge lengths $2^k$, with $k \leq K_{\text{SMRE}} := 5$.
If the origin of each square is shifted by $2^k$ in each direction the pattern is mapped onto itself. 
Only shifts by $2^i$ pixels in each direction with $i<k$ pose a mapping on a different pattern. 
The set $\tilde{\Omega}_k^i$ denotes all non-overlapping squares with an edge length $2^k$ which are shifted by $2^i$ pixels. 
Then construct the sets
\begin{align*}
    \Omega^i = \bigcap_{i\leq k \leq 5} \tilde{\Omega}_k^i. 
\end{align*}
The incomplete projection
\begin{align*}
    P_{\text{incomplete}} \left( x \right) = \prod_{i=0}^5 P_{\Omega^i}  x
\end{align*}
is computed as the projection on each $\Omega^i$ subsequently.
The resulting residual is still in the allowed set of $G$ in
Eq.~\eqref{eq2}, which preserves the statistical interpretation of the
result. 
\par
The projection on a set $\Omega^i$ is calculated in a single CUDA kernel (\autoref{lst:incomplete}). 
The kernel takes as input a pointer $e$ to the residual vector on the device, the dimension of the vector \lstinline{(ni, nj)}
and the offset \lstinline{(offseti, offsetj)} from where to generate the subsets.
The shifting is carried out as shown in \autoref{lst:incompletecall} by calling the kernel with different offsets~\lstinline{i}.
This increases the number of considered subsets.
The edge length $2^{\text{\lstinline{smin}}}$ of the smallest subsets can be specified to avoid unnecessary duplicate computations as a result of the shifting.
As in the kernel for the exact projection we make use of shared memory for the temporary arrays \lstinline|q|, \lstinline|s_1|, \lstinline|s_2|. 
The exact projection kernel uses 12 bytes of shared memory in case of single-precision and 24 bytes for double-precision per thread, the incomplete
kernel uses 32 bytes for single-precision per thread.
The shared memory is limited to 48KB per thread-block. For a reasonable implementation of the ICD we need the full 1024 threads per thread block rendering computations in double-precision infeasible on all existing generations of the CUDA architecture.



\begin{lstlisting}[language=c++, mathescape, morekeywords={__global__, __device__, __shared__, threadIdx},
escapebegin=\color{mygreen}, caption={Wrapper function around the incomplete Dykstra kernel}, label=lst:incompletecall]
template<typename T>
void ICD_handler() {
    int gridsi = info->width/32;
    int gridsj = info->height/32;
    dim3 grid(gridsi,gridsj);
    dim3 blocks(1024,1);
    for (int i = 0; i < 5; i++) {
        __incomplete_dykstra<T><<<grid,blocks>>>(info->e_device,
                                                 info->width,
                                                 info->height,
                                                 2^i, 2^i, i);
    }
}
\end{lstlisting}

\begin{lstlisting}[language=c++, mathescape, morekeywords={__global__, __device__, __shared__, threadIdx},
escapebegin=\color{mygreen}, caption={incomplete implementation of Dykstra's algorithm}, label=lst:incomplete]
// CUDA adapted incomplete Dykstra projection
// @param e pointer to the residual
// @param ni height of the image
// @param nj width of the image
// @param offseti offset in vertical direction
// @param offsetj offset in horizontal direction
// @param smin minimum subset size is $2^{\text{smin}}$

// `T` can be float or double, on current hardware only float is supported
template<typename T>
__global__ void
__incomplete_dykstra(T *e, const int ni, const int nj, const int offseti, const int offsetj, const int smin) {
    // Allocate temporary arrays q, s1, s2 in shared memory
    // ...
    
    // Set q = 0
    for (int s=5; s>=0; s--) {
        q[s*1024+threadIdx.x]=0;
    }
    
    // Temporary variables
    unsigned int is, js, idx;
    
    // Tolerance
    const T tol = TOLERANCE; //= 1e-3
    
    // Do parallel projections until convergence test passes
    T delta = 2.0*tol;
    while ( delta > tol*1024.0 ) {
        delta = 0;
        // Wait for all threads before starting the iteration
        __syncthreads();
        
        // In one threadblock we apply dykstra's algorithm to 
        // subsets with edge lengths 32, 16, 8, 4, 2 and 1.
        // Each thread processes one pixel.
        for (int s = 5; s >= smin; s--) {
            // Edge length of the subset = $2^s$
            int SubsetLength = pow_of_two(s);
            // Number of subsets in one threadblock = $2^{5-s}$
            int SubsetNum    = pow_of_two(5 - s);
            // Number of pixels in each subset = $2^{2\cdot s}$
            int SubsetSize   = pow_of_two(2*s);
            // Get Line in global image, assign to is
            // ...
            // Column in global image, assign to js
            // ...
            // For this iteration this thread is supposed
            // to process the pixel (is, js), the pixel
            // index idx in the global image is given by:
            idx = is*nj + js;
            
            // Fill shared memory with variables we use later
            s1[threadIdx.x] = e[idx] - q[s*1024+threadIdx.x];
            s2[threadIdx.x] = s1[threadIdx.x]*s2[threadIdx.x];
            // Wait for all threads
            __syncthreads();
            
            // Index to first pixel of my subset
            int SubsetStartIdx = (threadIdx.x/SubsetSize)*SubsetSize;
            
            // Sum over all pixels of one subset, write the result 
            // to s2[SubsetStartIdx]
            while (m <= pow_of_two(2*s-1)) {
                if (threadIdx.x - SubsetStartIdx + m < SubsetSize) {
                    s2[threadIdx.x] += s2[threadIdx.x + m];
                }
                m = m << 1; // m = m*2
                __syncthreads();
            }
            
            // $q = x_{r+1} - x_r$
            q[s*1024 + threadIdx.x] = -s1[threadIdx.x];
            
            // Eq. $\eqref{eq:dykstra:if}$ from Dykstra's algorithm
            // `cs` is a array allocated in CUDA constant memory
            if (cs[s]*s2[SubsetStartIdx] > 1.0) {
                s1[threadIdx.x] /= sqrt(cs[s]*s2[SubsetStartIdx]);
            }
            // Update $q$
            q[s*1024+threadIdx.x] += s1[threadIdx.x];
            // Calculate increment, mabs is our abs function
            delta+=mabs(e[idx] - s1[threadIdx.x]);
            
            // Update the estimate of residual
            e[idx] = s1[threadIdx.x];
            // Wait for all threads before next step
            __syncthreads();
        }
    }
}
\end{lstlisting}
\clearpage
\section{Results}
\label{sec:results}
The CUDA implementation of the new ICD variant is found to be 100 times faster than a serial implementation of Dykstra's algorithm and 10 times faster than the CUDA implementation of the exact version.
Yet, it reconstructs images nearly as well as the exact implementation, i.e. for the naked eye the results are indistinguishable.
Both implementations proved robust against noise and blur. 
The runtime performance is tested on synthetic images of different sizes.
Because of the enormous memory consumption of the exact variant of Dykstra's algorithm we had to vary the SMRE depth $K_{\text{SMRE}} $ for the different tests in order to keep the computations feasible. 
As an example of a real-life application we use the ICD as pre- and postprocessor for the SOFI algorithm which gives a further increase in resolution of up to 30\%.

\subsection{Test Data}
\label{sec:results:test_data}

\begin{figure}[p]
\centering
\hfill
\begin{subfigure}{0.47\textwidth}
 \includegraphics[width=\textwidth]{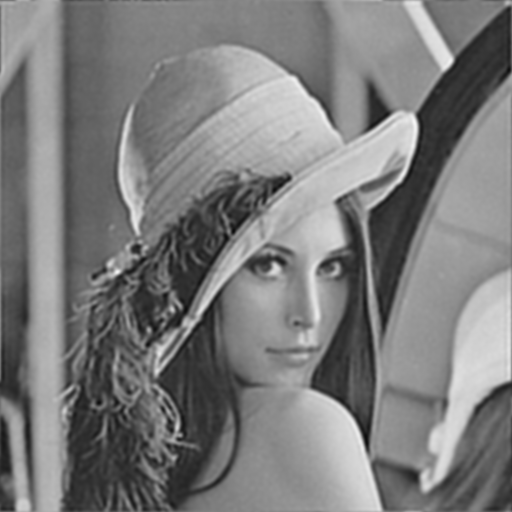}
 \caption{Test image, true signal.\newline\quad}
 \label{fig:lena:orig}
\end{subfigure}
\hfill
\begin{subfigure}{0.47\textwidth}
 \includegraphics[width=\textwidth]{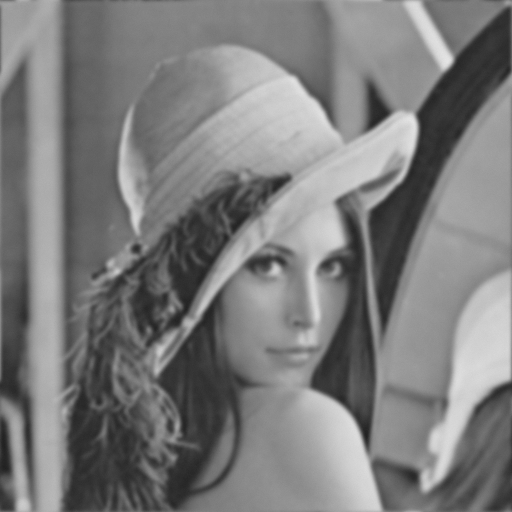}
 \caption{Test image with simulated blur and Gaussian noise of strength $\sigma = 1$.}
 \label{fig:lena:noise}
\end{subfigure}
\hfill\vfill\hfill
\begin{subfigure}{0.47\textwidth}
 \includegraphics[width=\textwidth]{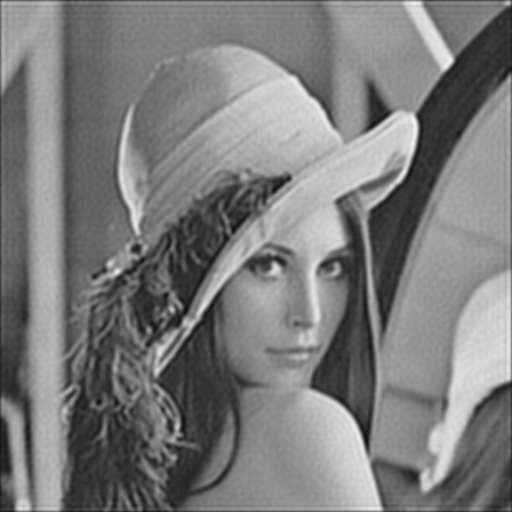}
 \caption{Reconstruction using exact implementation, noise strength $\sigma = 1$.}
 \label{fig:lena:exact}
\end{subfigure}
\hfill
\begin{subfigure}{0.47\textwidth}
 \includegraphics[width=\textwidth]{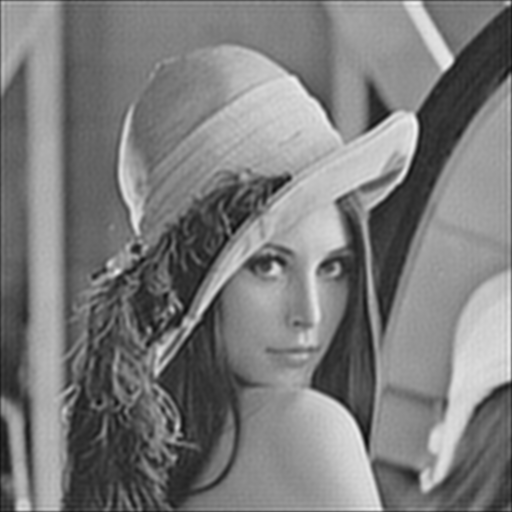}
 \caption{Reconstruction using incomplete implementation,  noise strength $\sigma = 1$.}
 \label{fig:lena:ICD-s1}
\end{subfigure}
\hfill\vfill\hfill
\begin{subfigure}{0.47\textwidth}
 \includegraphics[width=\textwidth]{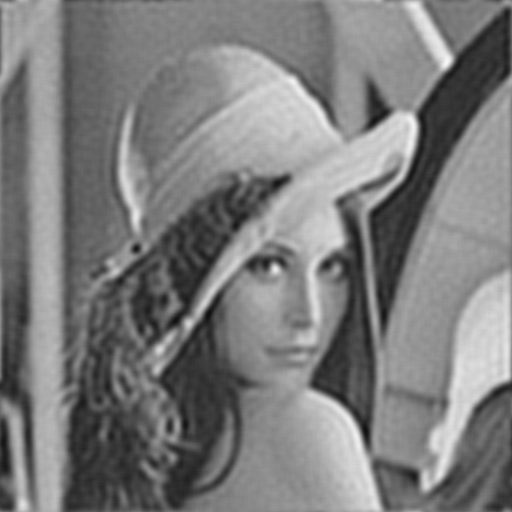}
 \caption{Reconstruction using incomplete implementation,  noise strength $\sigma = 3$.}
 \label{fig:lena:ICD-s3}
\end{subfigure}
\hfill
\begin{subfigure}{0.47\textwidth}
 \includegraphics[width=\textwidth]{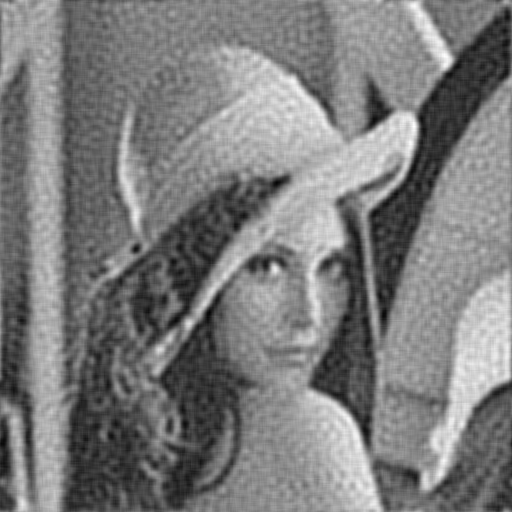}
 \caption{Reconstruction using incomplete implementation,  noise strength $\sigma = 10$.}
 \label{fig:lena:ICD-s10}
\end{subfigure}
\caption{Test of the algorithms on the Lena image, for details refer to Secs.~\ref{sec:results:test_data} and~\ref{sec:results:quality}. 
}
\label{fig:lena}
\end{figure}

To assess the quality of the estimator we use synthetic data  generated from the standard Lena image with simulated noise and blur.
For synthetic data the true signal is known and the quality
of the reconstruction algorithm can be estimated directly.
Figure~\ref{fig:lena:orig} shows the unperturbed standard Lena test image.
The simulated test image (Fig.~\ref{fig:lena:noise}) is generated by blurring the original image with a Gaussian kernel of width $\sigma_{\textrm{PSF}}=\SI{4}{px}$ (px = pixel) and
adding Gaussian noise with a standard deviation of $\sigma = 1$.
The noise strength $\sigma = 1$ corresponds to about 0.5\% of the maximum signal and is given in gray levels.
 We generated further test images with smaller signal to noise ratios, $\sigma=3$ and $\sigma=10$ (not shown), for comparison.
%


To assess the performance in real applications we study the combination of SMRE with SOFI (see Sec. \ref{section:sofi}) on 
 experimental data.
\par

\begin{figure}
\hfill
\centering
\begin{subfigure}{0.48\textwidth}
 \includegraphics[width=\textwidth]{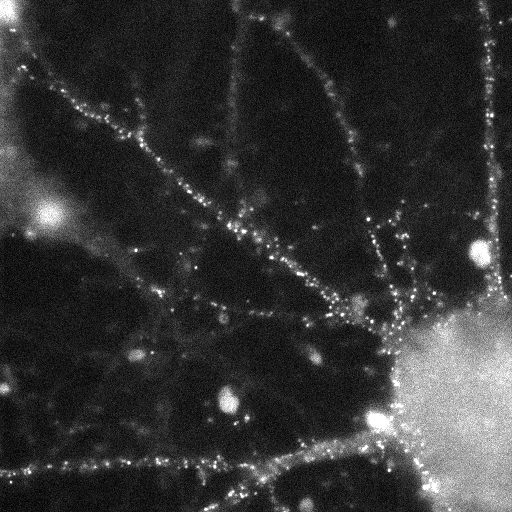}
 \subcaption{original frame}
 \label{fig:single:sofi}
\end{subfigure}
\hfill
\begin{subfigure}{0.48\textwidth}
 \includegraphics[width=\textwidth]{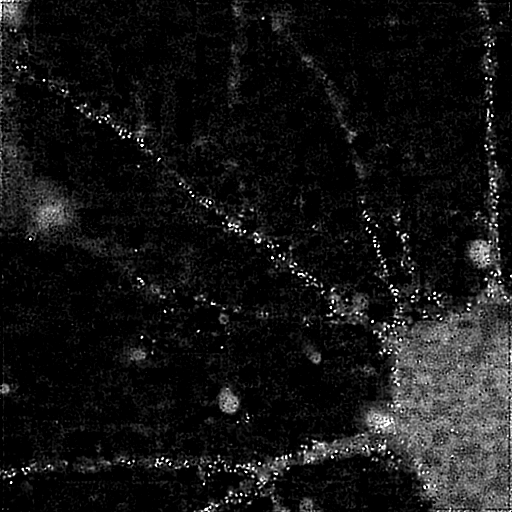}
 \subcaption{SMRE reconstruction}
 \label{fig:single:sofi:smre}
\end{subfigure}
\hfill
\caption{
Sample frame from the experimental time series of the hippocampal neurons \autocite{anja} \subref{fig:single:sofi}, and its SMRE reconstruction \subref{fig:single:sofi:smre}.
 }
 \label{fig:single}
\end{figure}

The final test of the exact and incomplete implementation of Dykstra's algorithm is done on optical widefield microscopy data, previously analyzed with the SOFI framework \autocite{anja} in order to study the details of intracellular trafficking and assembly
 of GABA-B neurotransmitter receptors in hippocampal neurons.
 The data set represents a time series of the temporal fluctuations of the fluorescence signal of a nerve cell labeled with quantum dots which attach to the receptors. 
 The fluctuations of the fluorescence stems from the blinking behavior of the quantum dots which is random and follows a power-law for the on- and off-times \autocite{qdotBlinkingPowerLaw}.
 The movie consists of 3091 frames. Besides the considerable amount of out-of-focus light due to the widefield illumination, each image shows the noise and blur typical for optical imaging. 
 A sample frame from the raw data is shown in Fig.~\ref{fig:single:sofi}.
The blur occurs due to the diffraction limit. 
The microscope records the real object convolved with the PSF.
We approximate the convolution kernel as Gaussian. In terms of the SMRE framework this is the measurement operator.


The images are typically acquired in the photon-limited regime. Hence one has to deal with Poissonian noise disturbing the image, i.e. to apply the Anscombe transformation before running the SMRE.

\subsection{Performance}
\begin{figure}[ht]
\centering
\hfill
\begin{subfigure}{0.48\textwidth}
 \includegraphics[width=\textwidth]{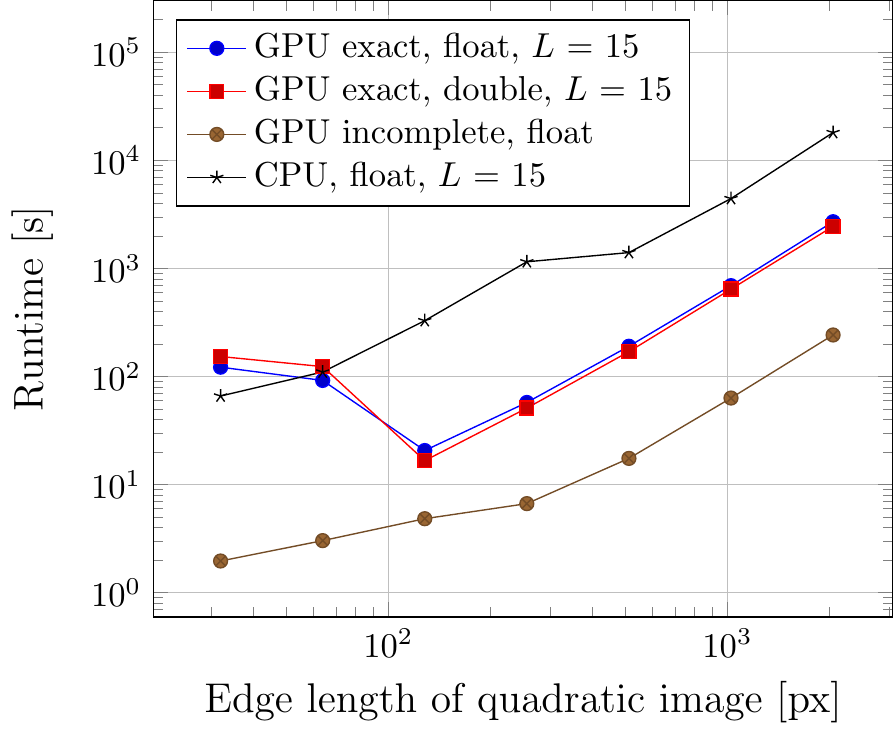}
 \subcaption{runtime}
 \label{fig:speed:run}
\end{subfigure}
\hfill
\begin{subfigure}{0.48\textwidth}
 \includegraphics[width=\textwidth]{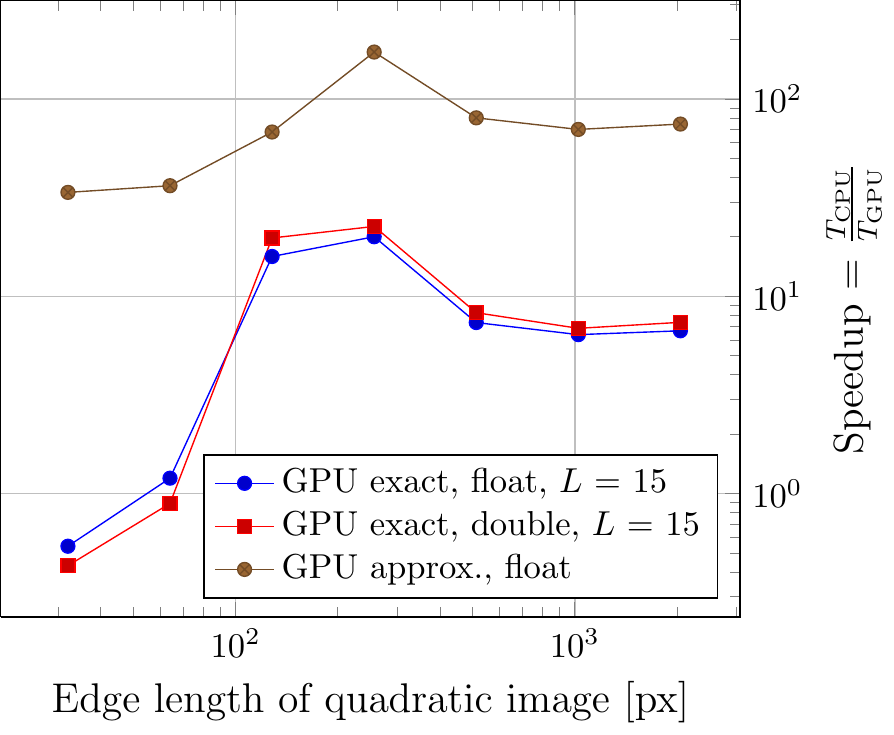}
 \subcaption{speedup}
 \label{fig:speed:up}
\end{subfigure}
\hfill
\caption{Runtime \subref{fig:speed:run} and speedup \subref{fig:speed:up} for 100 ADMM iterations. The exact GPU variants and the CPU variants use a maximum multi-resolution scale $L_{\text{SMRE}}=15$. 
All runtimes given exclude the startup time.}
\label{fig:speed}
\end{figure}
Figure \ref{fig:speed:run} compares the runtime of 100 ADMM iterations (in each iteration step Dykstra's projection algorithm is run until convergence)
between a serial implementation on a CPU (Intel Xeon X5675) and the exact and incomplete implementation on a NVIDIA Tesla K20c GPU.
For all variants we used a fixed tolerance of $10^{-3}$ for the Dykstra algorithm.
For the exact GPU and the serial CPU implementation the runtime depends on the number of subsets. 
The number of subsets for the incomplete GPU implementation only depends on the edge length of the image $L_I$.
The plots do not include the startup time required for the initialization of the algorithms.
While the main factor of the startup time for the incomplete implementation is the time needed to load the image, the startup time for the exact and CPU variants depends on the number of subsets.
The latter algorithms need to preallocate memory for each $q_s$ array of every subset $s$.
For the plots the maximum multi-resolution scale is $L_{\textrm{SMRE}}=15$ to make the computations feasible. As explained in Sec.~\ref{sec:impl:exact} the exact method then requires 14~GB of memory when run in single precision. 
The runtimes of the different methods are considerably different. 
The incomplete method achieves an overall speedup of up to 100 over the single-threaded CPU implementation and is 10 times faster than the exact method.

\subsection{Estimator quality}
\label{sec:results:quality}

\begin{figure}[p]
\centering
\begin{subfigure}{1.0\textwidth}
\centering
\includegraphics[width=0.9\textwidth]{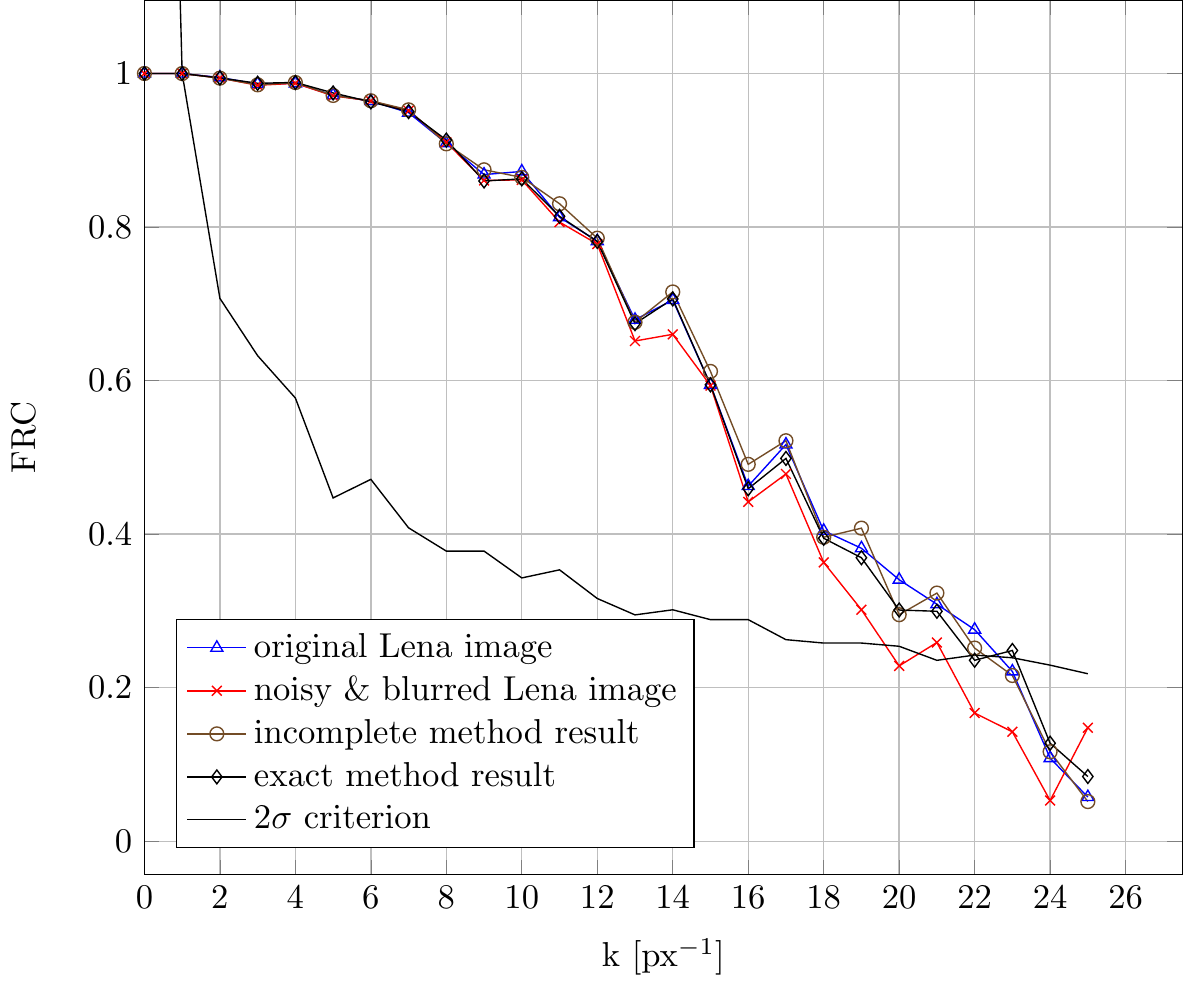}
\subcaption{FRC for simulated data and corresponding reconstruction with $\sigma_{\text{PSF}} = \SI{4}{px}$, $\sigma=1$. 
Note that both variants of Dykstra's algorithm produce nearly identical results.}
 \label{fig:frc:lena}
 \end{subfigure}
 \vfill
 \hfill
 \begin{subfigure}{0.485\textwidth}
 \includegraphics[width=\textwidth]{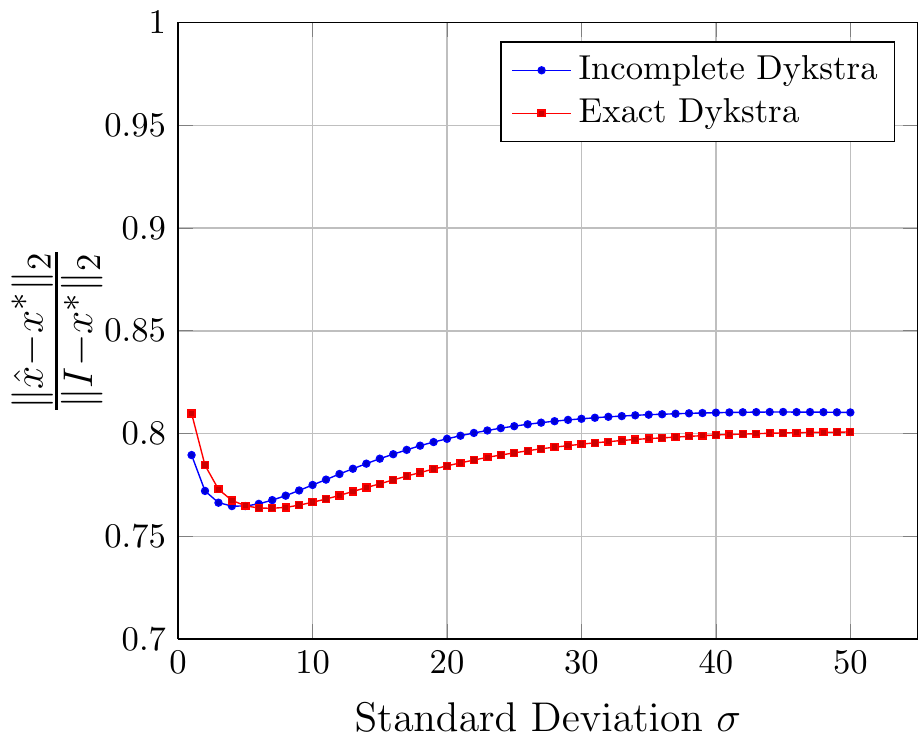}
 \subcaption{Robustness against noise level with fixed PSF radius of $\sigma_{\text{PSF}} = \SI{4}{px}$.}
 \label{fig:frc:lena_noise}
\end{subfigure}
\hfill
\begin{subfigure}{0.485\textwidth}
 \includegraphics[width=\textwidth]{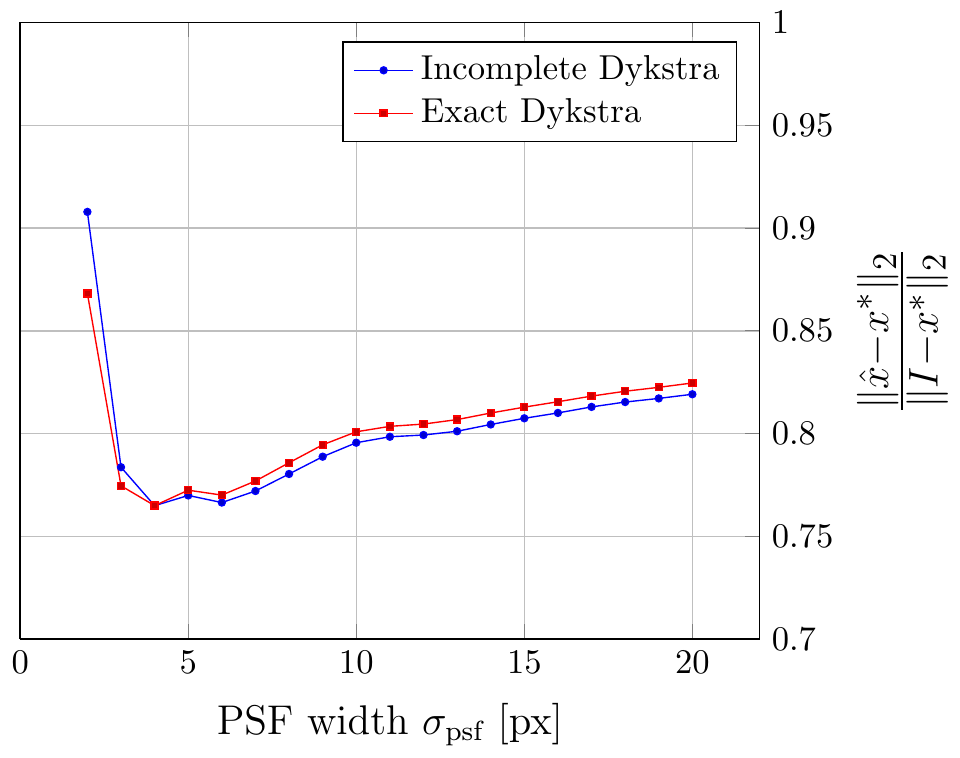}
 \subcaption{Robustness against PSF radius with fixed noise level $\sigma = 3$.}
 \label{fig:frc:lena_psf}
\end{subfigure}
\hfill
\caption{\subref{fig:frc:lena} Fourier ring correlation for the images from Fig. \ref{fig:lena}.
Robustness against noise strength $\sigma$ \subref{fig:frc:lena_noise} and PSF radius $\sigma_{\textrm{PSF}}$ \subref{fig:frc:lena_psf}, based on
the $L_2$-error $\|\hat x - x^*\|_2$ of the reconstruction $\hat{x}$ with respect to the true signal $x^*$ normalized  by the $L_2$-error $\| I - x^*\|_2$
of the simulated image $I$ with varying $\sigma$ (\subref{fig:frc:lena_noise}) 
or $\sigma_{\text{PSF}}$ (\subref{fig:frc:lena_psf}) for the Lena test image.
The exact implementation uses a maximum multi-resolution depth $L_{\text{SMRE}}=32$ in \subref{fig:frc:lena} and $L_{\text{SMRE}}=15$ in \subref{fig:frc:lena_noise} and \subref{fig:frc:lena_psf}.}
\label{fig:quality}
\end{figure}

Based on
the standard Lena test image, cf. Sec.~\ref{sec:results:test_data}, we analyze the quality of the statistical multi-resolution estimator.
%
Figure~\ref{fig:lena} summarizes  
the results from both implementations for  $\sigma=1$ (Figs.~\ref{fig:lena:exact} and~\ref{fig:lena:ICD-s1}) and the ICD results for  $\sigma=3$  (Fig.~\ref{fig:lena:ICD-s3}) and  $\sigma=10$ (Fig.~\ref{fig:lena:ICD-s10}).
All reconstructions use a maximum multi-resolution scale of $L_{\text{SMRE}}=32$.
The reconstructed image obtained from the exact implementation of Dykstra's 
algorithm (Fig.~\ref{fig:lena:exact}) is visually almost indistinguishable from the result obtained from the ICD variant  (Fig.~\ref{fig:lena:ICD-s1}), which is quantitatively confirmed by the FRC, cf. Fig.~\ref{fig:frc:lena}.

The FRC as described in Sec.~\ref{sec:frc} provides a quantitative measure of
the resolution.
Figure~\ref{fig:frc:lena} shows the FRC of the Lena images displayed in Fig.~\ref{fig:lena}. 
Visual inspection and the FRC indicate that the SMRE methods improve the image resolution and that the results of the
exact and incomplete implementation are nearly identical.

An important aspect of an estimator is robustness against noise strength and PSF radius.
Typically estimators tend to decrease in quality with increasing noise strength and PSF radius.
This behavior is investigated numerically in Fig.~\ref{fig:frc:lena_noise} and Fig.~\ref{fig:frc:lena_psf}, where
the robustness against noise and PSF radius is plotted as $L_2$-error $\|\hat x - x^*\|_2$ of the reconstruction $\hat{x}$ with respect to the true signal $x^*$, normalized 
by the error~$\| I - x^*\|_2$ of the simulated/measured image $I$. 
Values below $1$ indicate an improvement in quality of the reconstruction over the measurement. 
 Instabilities in the estimator become visible by
observing the proposed ratio.
Figure~\ref{fig:frc:lena_noise} shows the robustness against noise strength for fixed PSF radius $\sigma_{\text{PSF}} = \SI{4}{px}$.
The estimators are robust even for large noise.
For small noise is the incomplete estimator slightly less efficient. 
The plot indicates that the quality of the exact estimator is better
in terms of the $L_2$-distance to the true signal compared to the incomplete estimator.
The $L_2$-distance plot for the robustness against the PSF radius with a fixed noise level of $\sigma = 3$ (Fig.~\ref{fig:frc:lena_psf}), shows that the estimators are very robust against the PSF radius.
Here the exact and incomplete estimators are nearly identical in terms of the $L_2$-distance to the true signal.
The image quality improvement is about 20\% for both images in terms of the $L_2$-norm.

\subsection{SOFI results}
%


 \begin{figure}[htbp]
  \begin{center}
  \includegraphics[width=1.0\linewidth]{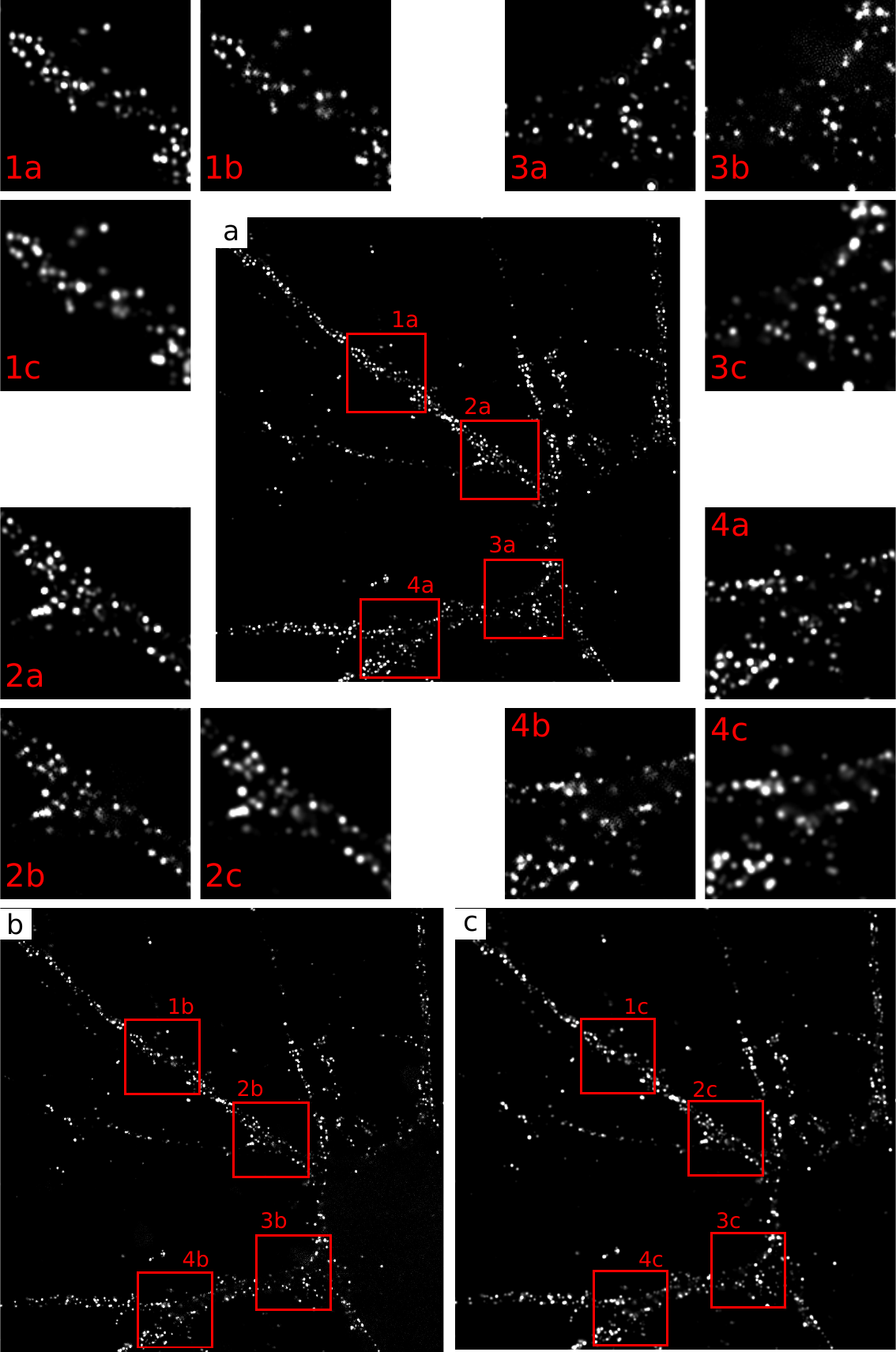}
  \caption{\textbf{(a)} SOFI second order with SMRE as preprocessor 
  on each raw image, \textbf{(b)} SMRE applied as postprocessor on the final SOFI second order image, \textbf{(c)} SOFI second order.
   The labels of the enlarged insets correspond to the region and reconstruction method respectively. As SMRE method we use ICD.}
  \label{fig:sofi2:comparison}
  \end{center}
 \end{figure}
 \begin{figure}[htbp]
 \centering
 \begin{subfigure}{0.7\textwidth}
  \includegraphics[width=\textwidth]{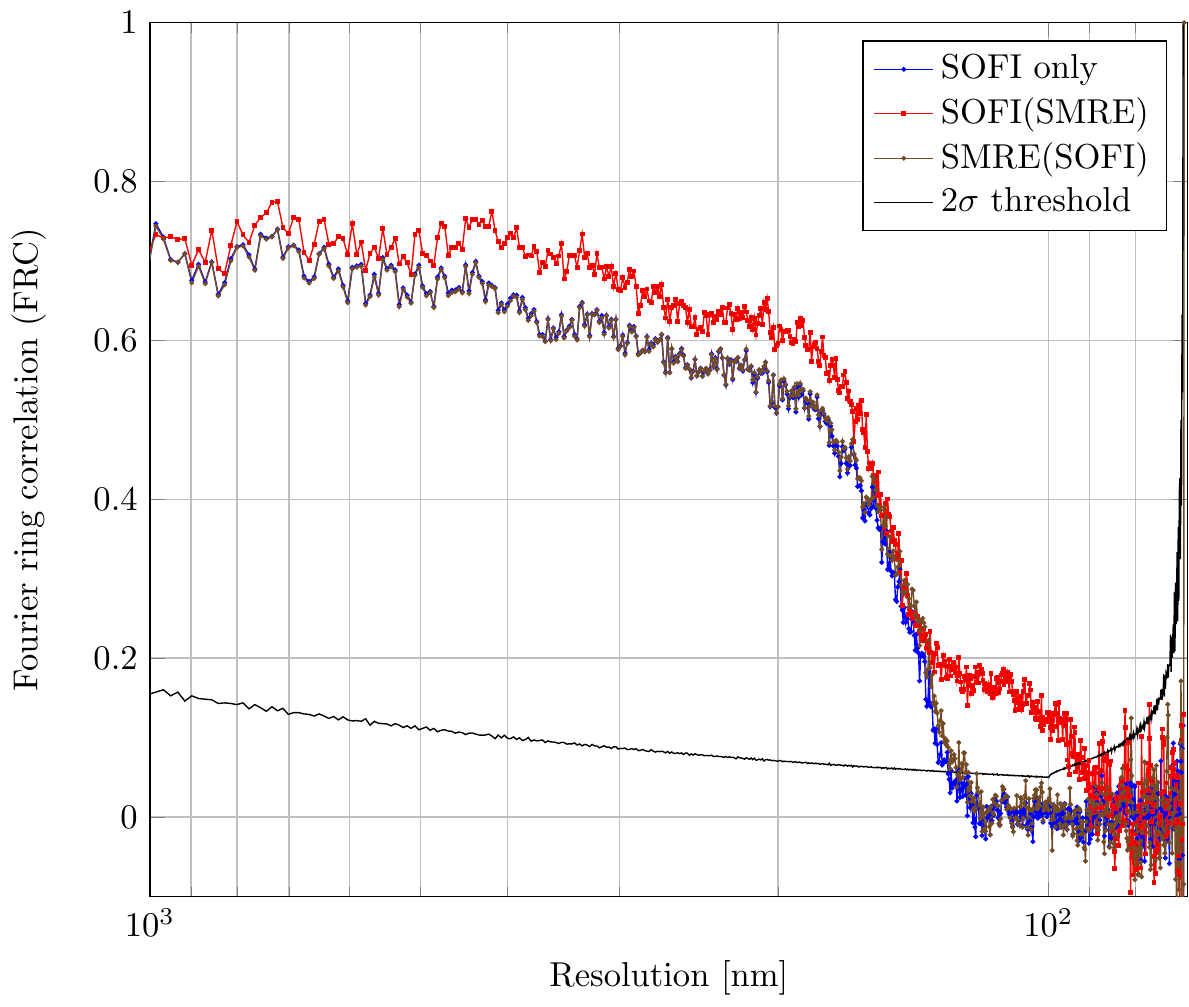}
  \subcaption{Mean FRC:
   The resolution is \SI{129}{nm} for the SOFI result, \SI{95}{nm} for SOFI with prior SMRE, and \SI{127}{nm} for SMRE with prior SOFI.
  }
  \label{fig:sofi2:frc}
 \end{subfigure}
 \begin{subfigure}{0.7\textwidth}
  \includegraphics[width=\textwidth]{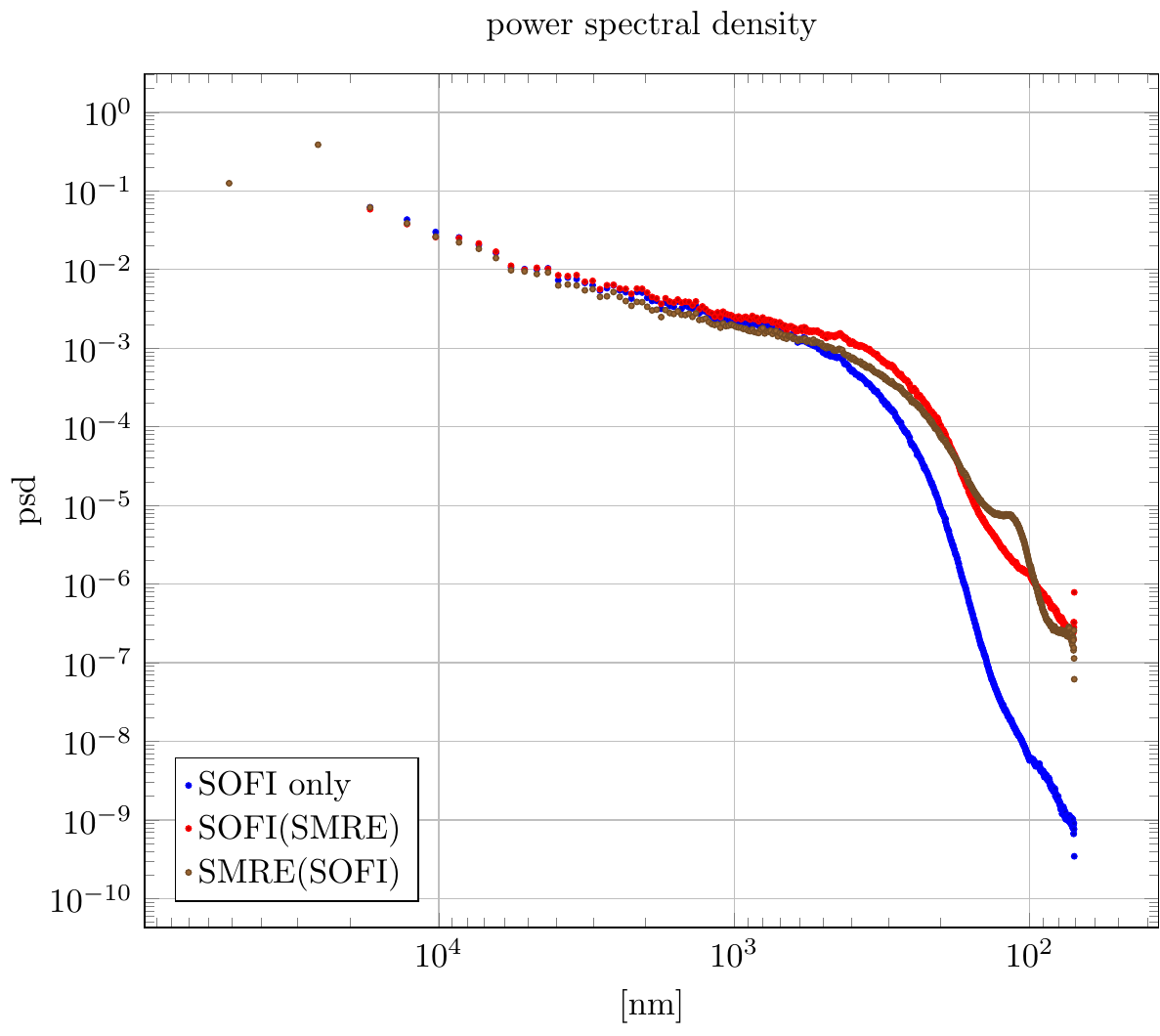}
  \subcaption{Normalized power spectral densities}
  \label{fig:sofi2:psd}
 \end{subfigure}
 \caption{Fourier ring correlation and power spectral density for the images in Fig.~\ref{fig:sofi2:comparison}. The resolution in the result with prior (SOFI$_2$(SMRE)) or posterior deconvolution  (SMRE(SOFI$_2$)) is better compared to the resolution without SMRE. }
 \label{fig:sofi2}
 \end{figure}

Figure~\ref{fig:sofi2:comparison} shows a comparison of using SMRE as a pre- and postprocessor for SOFI and the SOFI image without SMRE on the experimental dataset described in Sec.~\ref{sec:results:test_data}. 
Because of the much better performance all results were calculated using the ICD variant of Dykstra's projection algorithm.
In fact, it is the only variant for which these computations are feasible.
Due to the structure of the expected signal we use the generic $L_2$ norm as regularization.

In the following we denote by SOFI$_2$(ICD) the usage of the ICD variant of the SMRE as a preprocessor on each frame and by ICD(SOFI$_2$) the usage of SMRE as a postprocessor on the final SOFI image of second order.
If we rather want to stress that we use SMRE at all, we write  SOFI$_2$(SMRE) or SMRE(SOFI$_2$), respectively.

The comparison of SOFI$_2$(ICD) with ICD(SOFI$_2$) and SOFI without pre- or postprocessor in Fig.~\ref{fig:sofi2:comparison} shows a resolution improvement by using SMRE as a preprocessor for SOFI$_2$.
This leads to a greatly increased numerical effort, compared to SOFI$_2$ and ICD(SOFI$_2$). 
For ICD(SOFI$_2$) the workload shrinks from executing SMRE on several thousand frames to applying 
the SMRE concept to only one, i.e. the final, image. 
The FRC and the power spectral density (PSD) shown in Fig.~\ref{fig:sofi2} suggest a resolution improvement of about 35\% for SOFI$_2$(ICD). 
The resolution is estimated as intersection of the correlation curve with the $2\sigma$ threshold.
The resolution is \SI{129}{nm} for the SOFI$_2$ result, \SI{95}{nm} for SOFI$_2$ with prior SMRE, and \SI{127}{nm} for SMRE with prior SOFI$_2$.
%
%
The FRC and PSD curves in Fig.~\ref{fig:sofi2} are averages over several realizations of the reconstructions. The SOFI images were calculated from 1000 frames randomly chosen from the full data set.
Both methods indicate that the resolution in the result with prior SMRE 
is the highest.
\section{Discussion}
\label{sec:discussion}

We evaluated the performance of two different CUDA-implementations of Dykstra's algorithm.
The exact variant implements Dykstra's algorithm as defined in the mathematical literature with adaptions to the CUDA architecture while the incomplete variant only computes 
an approximation to Dykstra's projection by restricting the possible subsets.
In numerical comparisons this approximation 
seems to introduce only a small error. The result is nearly as good as the exact calculation while being much faster.

For the incomplete variant a rigorous proof of the convergence of the ADMM is still missing, but the numerical evidence for the convergence is encouraging. 

The exact implementation has significant disadvantages in terms of efficiency compared to the incomplete implementation.
While the algorithm generally achieves good efficiency and populates all the multiprocessors, it is limited by the
bandwidth of the PCIe bus, as data has to be constantly transferred between device and host. 
The memory transfers are necessary, because with increasing image size~$L_I$ and multi-resolution depth~$K_{\text{SMRE}}$ the total amount of memory required for storing all $q_s$ variables easily exceeds the global memory available on the device.
In that sense, the exact method cannot be implemented in a GPU-only fashion.

%

The usability of our SMRE approach in cooperation with super-resolution optical fluctuation imaging was shown with different datasets. 
Quite similar SMRE methods were deployed on images obtained from STED microscopy \autocite{hell, anscombe}. 

However, using SMRE for large-scale, high-throughput applications, e.g. as preprocessor in the SOFI method, is only feasible with the incomplete variant of Dykstra's algorithm which we have introduced in this paper.
The resolution improvements are in line with those obtained from Fourier-reweighted SOFI images where the deconvolution is done in frequency space.

The lesson to be learned from the success of the ICD is that the most crucial feature is the way the set of possible subsets is sampled.
A similar problem arises in Monte Carlo simulations of e.g. protein folding or multi-dimensional spin systems.
Typically the number of possible states of the system grows (for practical purposes) exponentially with the system size and thus is impossible to be completely enumerated.
Instead Monte Carlo methods rely on a clever sampling of the state space by using the appropriate probability distribution, which depends on the chosen algorithm. For Metropolis-Hastings \autocite{MC1953} it is the Boltzmann distribution and for the Wang-Landau \autocite{PhysRevLett.86.2050} or umbrella sampling it is the density of states, although it has to be constructed iteratively during the course of the simulation.
With that in mind the positive results of the incomplete variant of Dykstra's algorithm is much less surprising.
Given the fact the we only used the most obvious choice of a CUDA-friendly restriction of the shapes of the subsets we expect further speedups just by revisiting the way we have partitioned an image.

 There are certainly other aspects we still have to address in order to get a complete picture of the advantages and disadvantages of our statistical multi-resolution estimator and which we either had to omit or could touch only briefly in this paper.
 From a technical point of view one still has to investigate the influence of the choice and maximum size of the subsets 
 and the relationship of the termination criteria of the ADMM and Dykstra iterations to the quality of the reconstructed image.
 The SOFI method is only one way of super-resoluton microscopy and is mostly applied to widefield images.
 It is certainly of interest whether SMRE can be combined with other microscopy techniques, especially the parallel array microscope (PAM) \autocite{JMI:JMI945, doi:10.1117/12.879611}, which provides high-speed, high-throughput confocal imaging.
The SMRE algorithm is not restricted to 2D images. Therefore, a very interesting extension of the method would be to use 3D subsets for true 3D reconstructions.
Beyond all of these technical issues we would also like to test the performance of SMRE on other physical problems. For instance, deconvolution is an important topic in the context of fluorescence lifetime microscopy \autocite{JMI:JMI917} which is highly important for imaging ``single molecules in action''. The fluorescence lifetime crucially depends on the details of the chemical environment of a molecule and thus is able to track subtle interactions, for instance among proteins.

 \section*{Acknowledgments}
 
 We thank the Enderlein group from the III. Institute of Physics of the University of G\"ottingen for granting us access to their SOFI data and many fruitful discussions. 
 In particular we enjoyed an intensive collaboration with Anja Huss and Simon Stein.
 Most of the computations were done in the Institute of Numerical and Applied Mathematics of the University of G\"ottingen. Without the encouraging support by Gert Lube and Jochen Schulz this would not have been possible.
\clearpage
\printbibliography

\newpage
\appendix

\section{CUDA arch}
\label{sec:appendix-cuda-arch}

\subsection{CUDA programming model}

For the convenience of the reader we first briefly recall the CUDA-specific terms and the programming model.
CUDA and OpenCL allow to harness the compute power of modern GPUs for entirely non-graphical purposes.
The concepts are illustrated in Fig.~\ref{fig:cuda}.
GPUs in their current design offer thousands \autocite{Kepler} of independent compute cores, each executing a thread and each with its own dedicated amount of memory, the registers.
To distinguish the CPU, running the application, and the GPU executing data-parallel, compute-intensive tasks the notion of \textit{host} (CPU) and \textit{device} (GPU) was introduced.
Depending on the exact architecture a varying number of cores on the GPU forms a \textit{streaming multiprocessor} (SM).
The SM manages memory requests and issues instructions for threads. 
The threads are addressed in groups of 32 threads, which is called a \textit{warp}. 
In principle all threads of a warp execute the same instruction. 
Multiple warps form a thread block. 
A thread block resides on a single SM. 
Threads of the same thread block can exchange data via the \textit{shared memory} of the SM.
In order to process all of the input data, e.g. all pixels of an image, many thread blocks have to be started forming a so called \textit{grid}.
Data exchange between different thread blocks in a grid has to be done via the global device memory i.e. the VRAM. 
This communication inflicts already some latency which is hidden by a high level concurrency ("high occupancy") due to starting thousands of or more threads.
The bottleneck of many CUDA applications is the PCIe connection between host and device since it limits the memory transfer. This becomes an even bigger problem if the compute node has more than one GPU.

CUDA can be considered as a proprietary extension of the C programming language to GPGPU computing by adding the necessary keywords to handle data-parallel multithreading.
The API offers a way to call a kernel with specified thread block and grid dimensions which is executed on the GPU. 
Basically, a kernel is an ordinary function identified by the qualifier \lstinline|__global__|. 
In the simplest case each instance of a kernel processes an independent data element per computation.
The data element corresponding to a given thread, which has been started upon kernel launch, can be calculated from the thread's position in the grid of threads by means of the built-in variables \lstinline|ThreadIdx|, \lstinline|BlockDim|, \lstinline|BlockIdx| and \lstinline|GridDim|.
If threads have to frequently access the same data elements or exchange data with other threads, the shared memory of the SM can be used as a manually managed cache. 
Access to the shared memory is coalesced if all threads of a warp target the same cache line. Otherwise the access is serialized in as many accesses as there are cache lines addressed.
Shared memory is allocated by prepending the \lstinline|__shared__| qualifier to a variable in the scope of the kernel.
Synchronization among threads of a thread block is done by the \lstinline|__syncthreads()| method.
This is especially necessary for data exchange via the shared memory.
 
The kernel call itself runs asynchronous to the host part of the program, this means the next instruction after the kernel launch is immediately executed.
If this is e.g. a memory transfer of the results of the kernel just started, it will probably not yield the corrected results. 
Thus the host has to wait until the kernel finishes, this is achieved by using the \lstinline|cudaDeviceSynchronize()| directive.

In some situations, as we will show in the following sections, we can make use of this asynchronous behavior to achieve an optimal occupancy of the device.
With the advent of the Kepler architecture \autocite{Kepler}  Nvidia introduced an improved mechanism, called Hyper-Q, to handle concurrent execution of kernels and memory transfer via the use of \textit{streams}.
A CUDA stream is a sequence of operations that execute in issue-order on the GPU. 
Operations in different streams may run concurrently, before Hyper-Q the concurrency was limited as the CUDA streams were multiplexed into a single hardware work queue.
The Kepler architecture provides 32 work queues with no inter-stream dependencies. For a recent example of using Hyper-Q see \autocite{HyperQ}.

\section{Example parameters file}
\begin{lstlisting}[language=Gnuplot, keywordstyle=\color{black}]
# Listing of Parameters
# ---------------------
subsection input data
  # path to the .tif image
  set image          =
  # enforces first constraint
  set alpha1         = 1.2
  # enforces second constraint
  set alpha2         = 0.12
  # Estimate of gaussian noise standard deviation. 
  # If simulate, gaussian noise will be added to the image.
  set gaussian noise = 0
  # intensity of regularisation
  set regularization = 1.0
  # stabilises first constraint
  set rho1           = 6.192
  # stabilises second constraint
  set rho2           = 1.8
  # PSF spread
  set sigma          = 3
end


subsection output
  # where should we put the output image? Will be a tiff image
  set output image = control.tif
  # save preliminary results of the output image
  set control      = false
end


subsection program flow control
  # largest patch edge length if not using small dykstra in approximation
  set MRdepth            = 15
  # do a small dykstra in approximation
  set approx             = false
  # Set maximum number of iterations
  set maximum iterations = 10000
  # Reporting progress in intervals of ... Iterations
  set report interval    = 1
  # Finish when |x_r - x_{r-1}| < tolerance
  set tolerance          = 1e-3
end


subsection simulate dataset from real image
  # If set to false the input is treated as real data, 
  # if true input will be treated as test image where 
  # blurring and noise are added
  set simulate  = true
  # If false simulated noise has a constant seed, 
  # if true the seed is taken from the clock
  set time seed = false
end
\end{lstlisting}

\end{document}